\documentclass[twocolumn]{aastex62}
\usepackage{enumerate}
\usepackage{natbib}
\usepackage{hyperref}
\usepackage{amsmath}

\received{January 1, 2018}
\revised{January 7, 2018}
\accepted{\today}
\submitjournal{ApJ}

\shorttitle{NGC 300 IFU feedback}
\shortauthors{McLeod et al.}

\begin{document}

\title{Stellar Feedback and Resolved Stellar IFU Spectroscopy in the nearby Spiral Galaxy NGC 300}

\defcitealias{mcleod18b}{MC19}

\correspondingauthor{Anna F. McLeod, NASA Hubble Fellow}
\email{anna.mcleod@berkeley.edu, anna.mcleod@ttu.edu}

\author{Anna F. McLeod}
\affiliation{Department of Astronomy, University of California Berkeley, Berkeley, CA 94720, USA}
\affiliation{Department of Physics \& Astronomy, Texas Tech University, PO Box 41051, Lubbock, TX 79409, USA}

\author{J. M. Diederik Kruijssen}
\affiliation{Astronomisches Rechen-Institut, Zentrum f{\"u}r Astronomie der Universit{\"a}t Heidelberg, M{\"o}nchhofstra\ss e 12-14, D-69120 Heidelberg, Germany}

\author{Daniel R. Weisz}
\affil{Department of Astronomy, University of California Berkeley, Berkeley, CA 94720, USA}

\author{Peter Zeidler}
\affiliation{Department of Physics and Astronomy, Johns Hopkins University, Baltimore, MD 21218, USA}
\affiliation{Space Telescope Science Institute, 3700 San Martin Drive, Baltimore, MD 21218, USA}

\author{Andreas Schruba}
\affiliation{Max-Planck-Institut f{\"u}r extraterrestrische Physik, Giessenbachstra\ss e 1, D-85748 Garching}

\author{Julianne J. Dalcanton}
\affil{Department of Astronomy, University of Washington, Box 351580, Seattle, WA 98195, USA}

\author{Steven N. Longmore}
\affil{Astrophysics Research Institute, Liverpool John Moores University, Liverpool, L3 5RF, UK}

\author{M{\'e}lanie Chevance}
\affiliation{Astronomisches Rechen-Institut, Zentrum f{\"u}r Astronomie der Universit{\"a}t Heidelberg, M{\"o}nchhofstra\ss e 12-14, D-69120 Heidelberg, Germany}

\author{Christopher M. Faesi}
\affiliation{Max Planck Institute for Astronomy, K{\"o}nigstuhl 17, 69117 Heidelberg, Germany}
\affiliation{Department of Astronomy, University of Massachusetts - Amherst, Amherst, MA 01003, USA}

\author{Nell Byler}
\affiliation{Research School of Astronomy and Astrophysics, The Australian National University, ACT, Australia}
\affiliation{ARC Centre of Excellence for All Sky Astrophysics in 3 Dimensions (ASTRO 3D), Australia}



\begin{abstract}

We present MUSE Integral Field Unit (IFU) observations of five individual HII regions in two giant star-forming complexes in the low-metallicity, nearby dwarf spiral galaxy NGC 300.  In combination with high spatial resolution Hubble Space Telescope photometry, we demonstrate the extraction of stellar spectra and classification of individual stars from ground-based IFU data at the distance of 2 Mpc.  For the two star-forming complexes, in which no O-type stars had previously been identified, we find a total of 13 newly identified O-type stars, and 4 Wolf-Rayet stars (two already known sources and two Wolf-Rayet star candidates that this work now confirmed).  We use the derived massive stellar content to analyze the impact of stellar feedback on the HII regions.  As already found for HII regions in the Magellanic Clouds, the dynamics of the analyzed NGC 300 HII regions are dominated by a combination of the pressure of the ionized gas and stellar winds.  Moreover, we analyze the relation between the star formation rate and the pressure of the ionized gas as derived from small ($<100$ pc) scales, both quantities being systematically overestimated when derived on galactic scales. With the wealth of upcoming IFU instruments and programs, this study serves as a pathfinder for the systematic investigation of resolved stellar feedback in nearby galaxies, delivering the necessary analysis tools to enable massive stellar content and feedback studies sampling an unprecedented range of HII region properties across entire galaxies in the nearby Universe.

\end{abstract}



\section{Introduction}\label{sec:intro}

Feedback from massive ($>$ 8 M$_{\odot}$) stars is the main driver of secular evolution in galaxies with masses similar to or smaller than the Milky Way (MW).  Indeed, state-of-the-art numerical simulations of star-forming molecular clouds (e.g.~\citealt{bate09,dale14}) and galaxy evolution (e.g. \citealt{vogel14,schaye15}) cannot reproduce key observables (e.g.~the star formation rate vs.~stellar mass relation, \citealt{noeske07}) without accounting for stellar feedback. Therefore, feedback from massive stars represents one of the main uncertainties in numerical star formation and galaxy evolution studies.

Stellar feedback (e.g.~protostellar jets, ionizing radiation, stellar winds, supernovae, see \citealt{krumholz14a}) is known to operate on individual cloud scales ($\sim$ 10s of pc), and numerical simulations have shown that they can shape galaxy properties on global ($\sim$ kpc) scales (e.g.~\citealt{scannapieco12,hopkins13}). While there is a good \emph{qualitative} understanding of the physical mechanisms that connect these two scales, the field is lacking \emph{quantitative} observations that directly link the effect of feedback to individual massive stars (or massive stellar populations).

The dynamics of HII regions are governed by a combination of ionizing radiation, winds, and supernovae \citep{mckee84}, but disentangling the various feedback mechanisms, which often act simultaneously and possibly on relatively short timescales (e.g.~supernovae), is observationally challenging (e.g.\ \citealt{lopez14}).  Moreover, while feedback studies of a few individual, nearby (MW and Magellanic Clouds) HII regions exist (e.g. \citealt{smith08,pellegrini10,m16,pillars,mcleod18b}), due to observational limitations such as small fields-of-view and large distance uncertainties, most lack the simultaneous {\it detailed} knowledge of the gas (i.e.~properties and kinematics) and the feedback-driving massive stars (i.e.~number and spectral type) on which physical HII regions properties (e.g.~luminosity, temperature, expansion rate) critically depend. This observational characterization of both the sources and the effects of stellar feedback is absolutely critical to measure the fraction of feedback energy and momentum that couples to the surrounding medium \citep[sometimes referred to as the `feedback efficiency', e.g.][]{crain15,kruijssen18}. Observationally, this quantity can be traced by comparing the ionizing photon output of a young stellar population to the observed H$\alpha$ luminosity of the surrounding HII region.

In a recent study, \citet{mcleod18b} (henceforth referred to as \citetalias{mcleod18b}) showcased the capability of integral field units (IFUs) to spectroscopically identify and classify massive stars within HII regions, while simultaneously deriving feedback-related quantities. Using the optical IFU MUSE on the Very Large Telescope, \citetalias{mcleod18b} studied a total of $11$ individual HII regions in the Large Magellanic Cloud (LMC) in terms of their massive stellar content and the related effect of stellar feedback, finding that (i) the expansion of the HII regions is driven by winds and the pressure of the ionized gas, and (ii) feedback has a negative effect on star formation (i.e.~quenching).

Quantifying the influence of feedback from massive stars across a broad range of environments requires undertaking similar analyses in hundreds to thousands of resolved HII regions across a wide range of galactic environments (and a correspondingly large number of galaxies).  As an important step in this direction, we obtained a large IFU mosaic of the nearby spiral galaxy NGC 300 (for an excellent preliminary demonstration of the capabilities of MUSE in extracting stellar spectra in crowded fields in NGC 300, see \citealt{roth18}). With a favorable inclination angle, a distance of ${\sim} 2$~Mpc \citep{dalcanton09}, a well-studied population of HII regions and giant molecular clouds (GMCs; e.g.\ \citealt{deharveng88,faesi14,faesi18,kruijssen19}), and high spatial resolution, multi-band Hubble Space Telescope (HST) coverage \citep{dalcanton09}, NGC 300 is ideally suited to showcase the power of IFU spectroscopy for deriving feedback quantities and simultaneous detailed knowledge of the massive stars driving it over a wide range of HII region properties.  Our dedicated observing campaign (PI McLeod) uses MUSE to image the central $7\arcmin{\times}5\arcmin$ of the star-forming disk of NGC 300 with a 35-pointing mosaic, giving access to over 100 HII regions and their stellar content at a spatial resolution of ${\sim}10$~pc. This type of IFU analysis of resolved HII regions will continue to grow with other ongoing and upcoming efforts (e.g.~Physics at High Angular resolution in Nearby GalaxieS, PHANGS, first results shown in \citealt{kreckel18}; Star formation, Ionized Gas, and Nebular Abundances Legacy Survey with SITELLE, \citealt{rousseau19}).

In this paper, we preview our larger program and showcase the combined power of HST+MUSE observations of two giant HII region complexes hosting a total of $5$ individual HII regions.  For each HII region we derive integrated properties (e.g.~H$\alpha$ luminosities, oxygen abundances, electron densities and temperatures) and feedback-related pressure terms (i.e.~the direct radiation pressure, the pressure from stellar winds, and the ionized gas pressure; due to the lack of X-Ray emission from archival Chandra data, the relatively young ages of the analyzed regions, the lack of common optical supernova remnant tracers e.g. enhanced [SII] emission with respect to H$\alpha$ and the absence of known remnants associated with the analyzed regions as per \citealt{vucetic15}, we assume that supernova events have not yet occurred); moreover, we extract spectra of the massive, feedback-driving stars in the HII regions, and derive stellar atmospheric parameters by fitting PoWR model atmospheres \citep{hainich19} to the detected stellar absorption lines.  With the combined knowledge of the region properties and the massive stellar content, we analyze the feedback efficiency by comparing the combined photon output from the stars to the region luminosities.

As a first comparison with similarly-derived parameters, we compare our results to those described in \citetalias{mcleod18b}, and discuss the potential of the full 35-pointing mosaic in terms of obtaining not only integrated HII region properties (e.g.~luminosities, abundances, temperatures, densities, ionized gas kinematics) over entire nearby galaxies, but also to determine their feedback-driving stellar content in terms of O-type and Wolf-Rayet (WR) stars.  The methods and techniques will be widely applicable to similar, existing IFU data sets (e.g.~PHANGS/MUSE, \citealt{kreckel18}, SIGNALS/SITELLE, \citealt{rousseau19}) and instruments (e.g.~Keck/KCWI), as well as upcoming instruments and surveys (e.g.~ELT/HARMONI, SDSS-V/LVM).  These will yield the necessary statistics in terms of HII region numbers and properties needed to observationally quantify stellar feedback across the wide range of physical conditions found in nearby galaxies. 

This paper is organized as follows.  In Section \ref{sec:data}, we introduce the MUSE data set, discuss the data reduction, and briefly discuss the HST photometric catalog we used.  In Section \ref{sec:stars}, we describe the stellar identification and classification process. In Section \ref{sec:gas}, we discuss the integrated properties of the five HII regions and compare the results obtained here to previous LMC feedback studies as well as simulations and local starburst galaxies.  In Section \ref{sec:outlook}, we provide an outlook for upcoming nearby galaxy IFU surveys, and we conclude in Section \ref{sec:end}.

\section{Observations and data reduction} \label{sec:data}

\subsection{MUSE IFU data}\label{sec:musedata}
\begin{figure*}[ht!]
\gridline{\fig{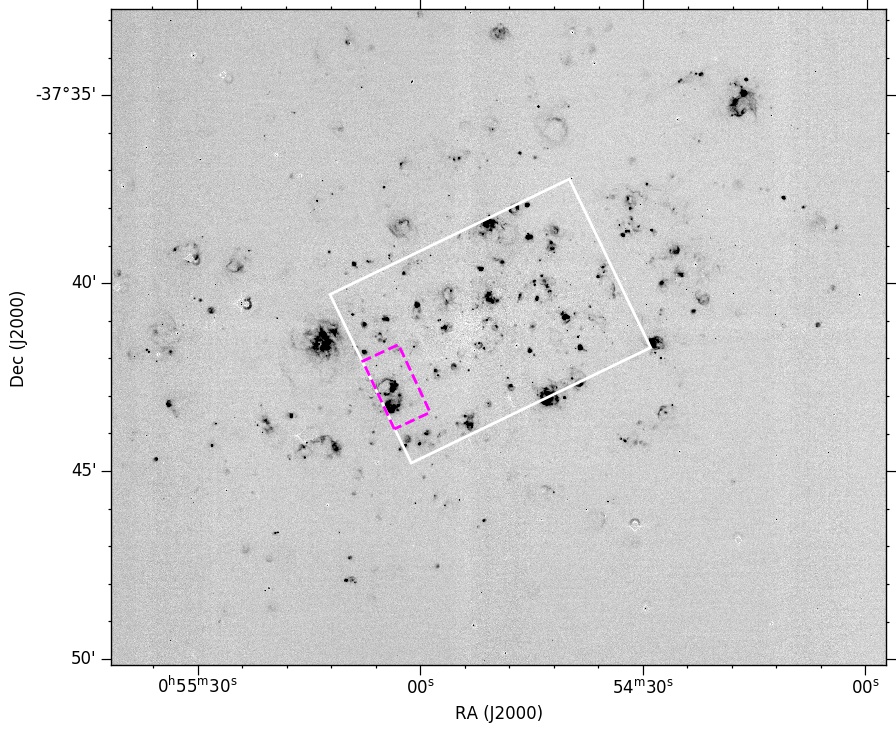}{0.7\textwidth}{(a)}}
\gridline{\fig{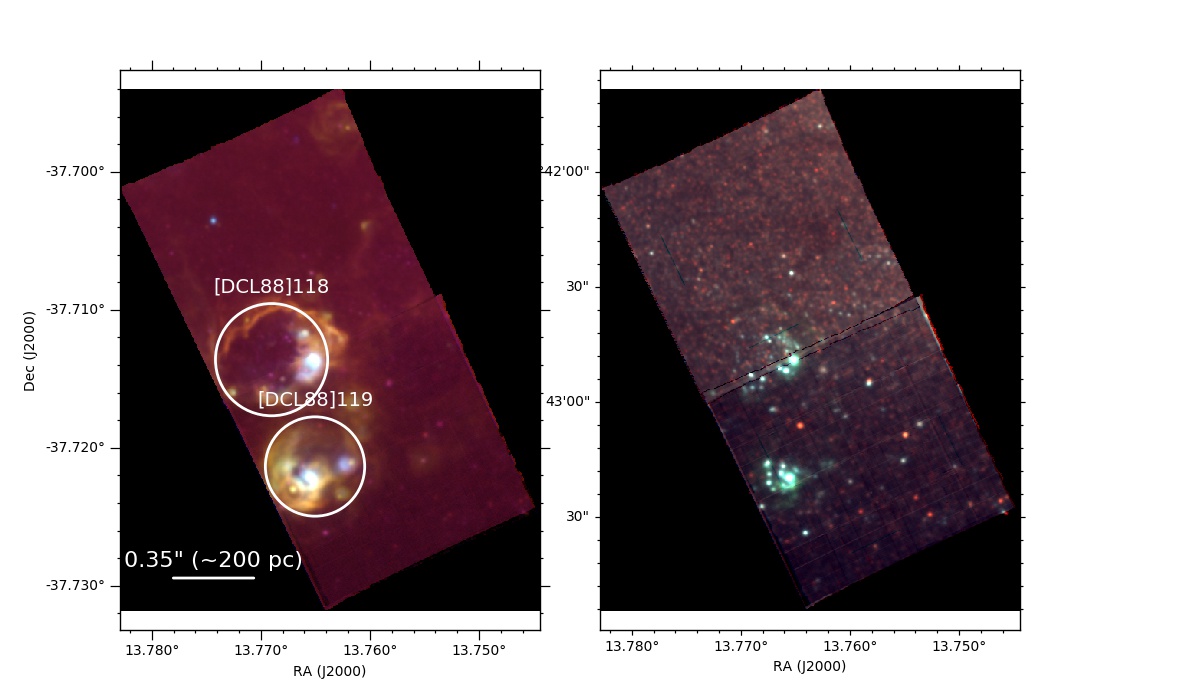}{0.9\textwidth}{(b)}}
\caption{(a) ESO/MPG WFI H$\alpha$ image of NGC 300 \citep{faesi14} showing the footprint of our $7\arcmin{\times}5\arcmin$ mosaic (solid white), and the two regions analyzed here (dashed magenta) (b) Three-color composites of the two HII region complexes. Left panel: ionized gas map traced by [SII]$\lambda$6717 (red), H$\alpha$ (green), and [OIII]$\lambda$5007 (blue), the white circles encompass $90\%$ of the H$\alpha$ flux in each region. Right panel: arbitrary $\mathit{RGB}$ (see text Section \ref{sec:musedata}) composite highlighting the stellar population. Some nebular emission is seen in the green filter, which encompasses the H$\alpha$ line.}
\label{rgb}
\end{figure*}

We used the MUSE instrument (Multi Unit Spectroscopic Explorer, \citealt{muse}), mounted on the Very Large Telescope, to obtain a $7\arcmin{\times}5\arcmin$ mosaic covering most of the star-forming disk of NGC 300.  The data were taken in the nominal wavelength range (${\sim} 4750{-}9350$~\AA) and in the instrument's wide-field mode ($1\arcmin{\times}1\arcmin$ per pointing), as part of the observing program \mbox{098.B-0193(A)} (PI McLeod).  These data were not taken with the Adaptive Optics system of MUSE, and seeing-limited angular resolutions in a range of 0\farcs7 - 1\farcs1 were achieved.  Each pointing was observed three times in a 90\degr-rotation dither pattern with an exposure time of $900$ seconds per rotation.  Here, we present a first analysis of the data set concerning two giant HII region complexes in the eastern part of NGC 300's star-forming disk, covered by two of the (in total) $35$ MUSE pointings of the program (shown in Fig.~\ref{rgb}), the details of which we list in Table~\ref{tab:pointings}. The two complexes host a total of five individual HII regions.  Following the nomenclature of \citet{deharveng88}, these are [DCL88]118A, [DCL88]118B, [DCL88]119A, [DCL88]119B, and [DCL88]119C. Henceforth, we refer to these HII region complexes as D118 and D119, and drop the [DCL88] before the regions' names for brevity. 

The two pointings were reduced in the \textsc{Esorex} environment using the MUSE pipeline \citep{pipeline} and the standard static calibration files.  For each observing block we used the available calibration files from the ESO archive for the relevant night. Standard star calibrations were performed as part of the instrument's pipeline, the relevant standard star observations\footnote{See the ESO/MUSE webpages for a list of standard stars used for the instrument.} (Feige110 and EG274 and for fields 1 and 2, respectively) were taken as part of the nightly MUSE calibration plan with an exposure time of 40 seconds.  The three exposures per pointing of the two data cubes were then combined into two single cubes with the built-in exposure combination recipes of the MUSE pipeline.

Our program does not include dedicated sky offsets, but with an exposure time of $900$ seconds the MUSE spectra are heavily contaminated by sky lines.  Atmospheric contamination increases towards the red part of the MUSE wavelength coverage, where many O and OH transitions occur.  This is particularly problematic in terms of spectral classification of e.g.~lower mass stars, as several crucial features (e.g.\ the Ca~II triplet) lie within that range.  Atmospheric contamination is also problematic in terms of ionized gas studies, as several nebular emission lines (e.g.\ the Balmer lines used for extinction measurements, the density-sensitive [SII]$\lambda$6717,31 doublet) are surrounded by and blended with sky lines.  

In the absence of dedicated sky observations, the MUSE pipeline relies on estimating the sky emission from a user-defined percentage of the darkest pixels in a given, specific data cube, and the so-created sky spectrum is then subtracted from all the spaxels (spectral pixels) in that specific cube.  For observations targeting regions with strong nebular emission this method is, unfortunately, not ideal, as the atmospheric contribution is not only due to O$_{2}$ and OH transitions, but also to emission from species that make up a nebular spectrum (i.e.~H$\alpha$, H$\beta$, and [OI]).  Using the sky creation and subtraction of the pipeline therefore leads to negative nebular fluxes, as the pipeline recognizes e.g.\ the Balmer and atomic oxygen lines as sky lines and subtracts them.  To avoid this, we proceed in removing the atmospheric sky lines in the following manner (a detailed description of this method is given in \citealt{zeidler19}):

\begin{enumerate}[(i)]
\item Standard sky creation: the appropriate pipeline recipe ({\it muse\_create\_sky}) is used to create a sky model by using a $0.5\%$ fraction of the input image to be considered as sky.
\item Sky continuum: the previous step produces an estimated sky continuum flux spectrum, which we now set to zero.  While no continuum subtraction is performed at this point, this step is necessary, as it ensures that the fluxes of the sky emission lines are estimated correctly, and that the input for the next step has the correct, pipeline-recognized header.
\item Zero-flux sky creation: {\it muse\_create\_sky} is called again with the sky continuum (set to zero, as described above) obtained from the previous step.
\item Sky line subtraction: the calibration file containing the list of sky lines, that comes as part of the pipeline's static calibration folder, is overwritten such that all but the atmospheric molecular lines (i.e.~O$_{2}$ and OH) have zero fluxes.  The science reduction recipe {\it muse\_scipost} is now run on the data cube together with the above created sky continuum and sky line calibrations. 
\end{enumerate}

\begin{figure*}[h]
\gridline{\fig{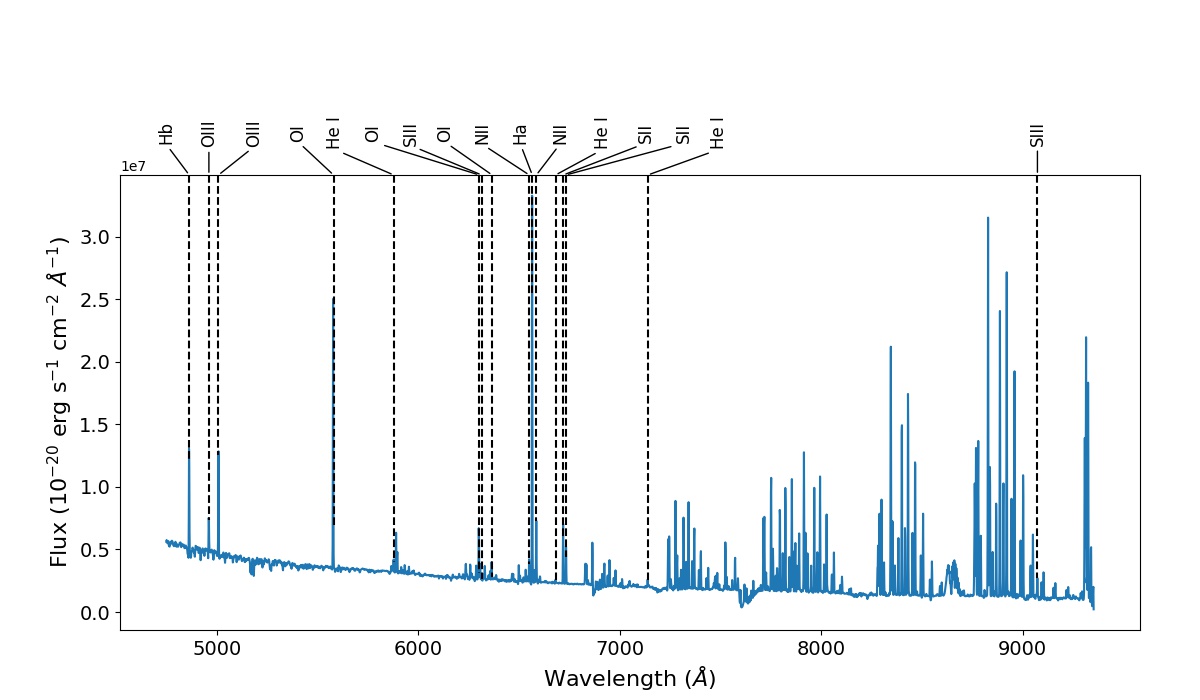}{0.9\textwidth}{}}
\vspace{-1.1cm}
\gridline{\fig{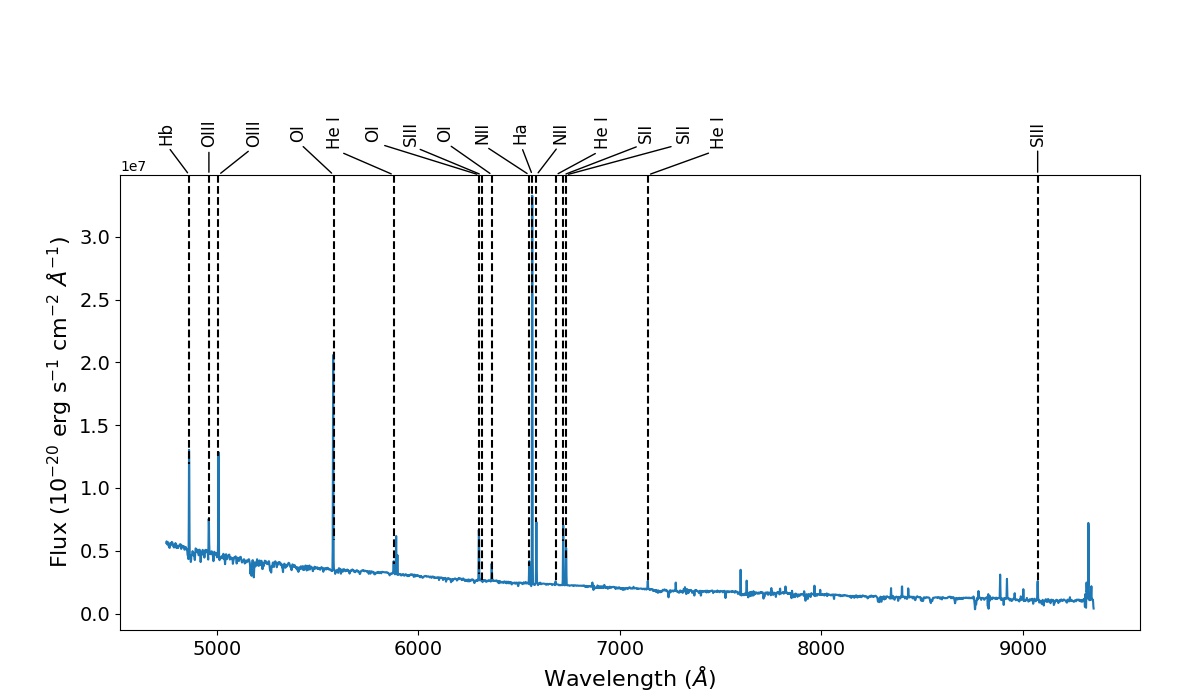}{0.9\textwidth}{}}
\caption{Integrated spectrum of Field 2 before (upper panel) and after (lower panel) sky subtraction.  The main nebular emission lines are indicated (unlabeled emission lines in the lower panel are residual atomic species of no direct relevance for this study).\label{fig:lines}}
\end{figure*}

The result of this procedure (which also includes a telluric correction) is shown in Fig.~\ref{fig:lines}, where the upper panel shows the integrated spectrum of Field~2 before the sky line subtraction, and the lower panel illustrates the output of the above-described method. The unlabeled emission lines in the lower panel are residual atomic species of no direct relevance (or contamination) for this study. 

Field~2 (see Table \ref{tab:pointings}) was taken during non-photometric conditions.  As a result, the three exposures of Field~2 are offset relative to another in flux.  We estimate the relative offsets by fitting the baseline of the mean spectrum of each individual cube of Field~2, and the obtained values are integrated into the MUSE pipeline during the exposure combination step which produces the final, combined data cube of Field~2.  

Moreover, due to lunar contamination, the flux offset between the three exposures is wavelength dependent, with an increasing offset towards the bluer wavelengths. In the data this is reflected by a flux increase towards bluer wavelengths.  Because the pipeline can account for constant but not for wavelength-dependent flux offsets between exposures, we do not correct for lunar contamination during data reduction.  Instead, this is taken into account when producing integrated emission line maps by performing a pixel-by-pixel continuum subtraction of the data cubes, as will be described in the following paragraph.

\begin{deluxetable*}{ccCrcr}
\tablecaption{MUSE IFU observations discussed in this paper. \label{tab:pointings}}
\tablecolumns{6}
\tablenum{1}
\tablewidth{0pt}
\tablehead{
\colhead{Field} & \colhead{Field center} & \colhead{Observation date} & \colhead{Seeing} & \colhead{t$_{\rm{exp}}$} & \colhead{\# of exposures} \\
\colhead{} & \colhead{(J2000)} & \colhead{(yyyy$-$mm$-$dd)} & \colhead{(arcsec)} & \colhead{(s)} & \colhead{}
}
\startdata
1 & 00:55:04.35 -37:42:19.8 & 2016{-}11{-}08 & 0\farcs70 & 900 & 3 \\
2 & 00:55:02.15 -37:43:13.5 & 2018{-}07{-}23 & $\sim$ 1\farcs06 & 900 & 3 \\
\enddata
\end{deluxetable*}

In this paper, we analyze both the stellar and the gaseous components of the five HII regions within the two large complexes.  Both components require absorption from the galactic stellar background to be accounted for. For the stellar spectra analyzed in this work this step is necessary to disentangle intrinsic stellar absorption features from those produced by the galactic background (this will be further discussed in Section~\ref{sec:stars}).  In terms of the nebular analysis, the stellar galactic background introduces absorption features in the Balmer lines, which are superimposed on the emission of nebular origin, and can lead to faulty line ratio values, e.g.~higher H$\alpha$/H$\beta$ ratios, used during extinction correction. We therefore proceed in performing a pixel-by-pixel continuum subtraction and background correction.  The continuum subtraction is performed by fitting and subtracting a fourth-order polynomial\footnote{The high polynomial order used for the continuum fit is motivated by the large fraction of continuum in each spectrum available thanks to precise user-defined wavelength ranges to be excluded from the fit available in the {\sc Pyspeckit} package used for this.} to each spaxel, followed by the subtraction of the local background. The former therefore removes both the lunar and the stellar continua simultaneously.  For the background subtraction, the mean of a 3-pixel-radius aperture (motivated by the PSF fit described in Section \ref{sec:stars}) containing pixels representative of the darkest $5\%$ of pixels (i.e.~without nebular emission and not coinciding with stellar sources) in each cube is used for the pixel-by-pixel background subtraction.  This results in a continuum- and background-subtracted data cube per field. These two cubes are then manually combined into a single, $2\arcmin{\times}1\arcmin$ data cube, which is used to derive the integrated (nebular) properties described in Section~\ref{sec:gas}. 

To validate the flux offset, stellar background, and continuum subtraction corrections performed prior to data analysis, we compare the MUSE H$\alpha$ fluxes obtained from the MUSE H$\alpha$ map by integrating over the two HII region complexes (white regions in Fig.~\ref{rgb}, corresponding to the radii encompassing 90\% of the H$\alpha$ flux, see Sec.~\ref{sec:params}) with fluxes obtained from the same apertures from ESO/WFI data\footnote{The WFI H$\alpha$ filter has a central wavelength of 6588.27 \AA~and a FWHM of 74.31 \AA, and therefore also covers the [NII] lines. As described in \citet{faesi14}, the [NII] contamination was removed and the continuum subtracted, hence the resulting H$\alpha$ map is comparable to the MUSE map which only cover the H$\alpha$ line.} \citep{faesi14} before and after the correction. Table~\ref{tab:muse_wfi}, showing good agreement between the WFI and corrected MUSE fluxes.

\begin{deluxetable}{cccc}
\tablecaption{Comparison of the H$\alpha$ fluxes obtained from the MUSE data set prior (column~2) and post (column~3) flux offset correction (see Section~\ref{sec:musedata}) with archival ESO/WFI H$\alpha$ fluxes \citep{faesi14}. WFI fluxes are obtained by performing aperture photometry on the circular regions shown in Fig.~\ref{rgb}. All fluxes are in units of 10$^{-14}$ erg s$^{-1}$ cm$^{-2}$ and are pre-extinction (MW + intrinsic) correction. \label{tab:muse_wfi}}
\tablecolumns{4}
\tablenum{2}
\tablewidth{0pt}
\tablehead{
\colhead{Region} & \colhead{F$_{\mathrm{MUSE}}$} & \colhead{F$_{\mathrm{MUSE, corr}}$} & \colhead{F$_{\mathrm{WFI}}$} 
}
\startdata
{[DCL89]}D118 & 50.61$\pm$0.01 & 58.68$\pm$0.02 & 60.84$\pm$0.06 \\
{[DCL89]}D119 & 75.47$\pm$0.01 & 106.26$\pm$0.01 & 105.69$\pm$0.07 \\
\enddata
\end{deluxetable}

To show the spatial distribution of the ionized gas in relation to the ionizing, massive stars, we produce emission line maps of the main nebular lines (by collapsing the cubes $\pm$3~\AA\ around the central, redshifted, wavelength of each line, i.e.~H$\beta$, [OIII]$\lambda\lambda$4959,5007, [NII]$\lambda\lambda$6548,84, H$\alpha$, [SII]$\lambda\lambda$6717,31, [OII]$\lambda\lambda$7320,30, [SIII]$\lambda$9068). We also produced maps in three custom filters, which simply divide the spectral axis of the cube into three arbitrary ranges. For simplicity, we call these $B$, $G$ and $R$ ($4750{-}6283$~\AA, $6284{-}8717$~\AA, and $8718{-}9350$~\AA, respectively).  The resulting three sub-cubes are then collapsed along the spectral axis to obtain images in each filter.  This is shown in Fig.~\ref{rgb}, as a nebular three-color composite image (the left panel shows r = [SII]$\lambda$6717, g = H$\alpha$, b = [OIII]$\lambda$5007), and the corresponding $\mathit{RGB}$ composite in the right panel).

\subsection{HST photometry}\label{sec:hst}
As will be described in Section~\ref{sec:stars}, we use high angular resolution HST photometry catalogs to de-blend and resolve single stars and then extract their spectra from the ground-based MUSE data.  NGC 300 was covered by the ACS Nearby Galaxy Survey \citep[ANGST;][]{dalcanton09} in three broadband photometric filters, i.e.\ F435W, F606W, and F814W. Additionally, ANGST combined these observations with archival HST data of NGC 300 \citep{butler04,rizzi06}, including F555W observations.

For each observed field, ANGST provides publicly available star catalogs, which contain the photometry for each object classified as a star and with $S/N>4$ (we refer to \citealt{dalcanton09} for details on photometry measurements and point source classification).  ANGST also provides ``good star" catalogs, in which additional sharpness and crowding cuts have been applied.  Because of these criteria, the good star catalog is not ideal when analyzing star clusters where stars are crowded by definition, and we therefore use the above-mentioned star files in this paper, rather than the ``good star" catalogs. 

From the ANGST catalog covering the two HII region complexes (and containing a total of 373,677 sources), we extract stars within 22\arcsec ($\sim$ 215 pc) of the central coordinates of the HII region complexes.  Given the angular sizes of D118 and D119 (which have radii of $\sim$ 14.6\arcsec and 13\arcsec, respectively, see Fig.~\ref{rgb}), $22\arcsec$ encompasses the cluster stars as well as field stars that are representative of the stars in the region surrounding each star cluster.  These more isolated field stars are necessary to compute the PSF, as will be described in Section~\ref{sec:stars}.  We further limit the HST catalog by selecting only bright sources, i.e.\ $m_{\rm{F814W}}<22$, corresponding to the $I$~band limiting magnitude of our MUSE observations, having SNR $>$ 10 in both filters (F435W and F814W). This results in two catalogs with $76$ and $69$ stars for D118 and D119, respectively. Typical photometric errors errors of ANGST magnitudes are of the order $\sim 10^{-3}$.

\begin{figure*}
\gridline{\fig{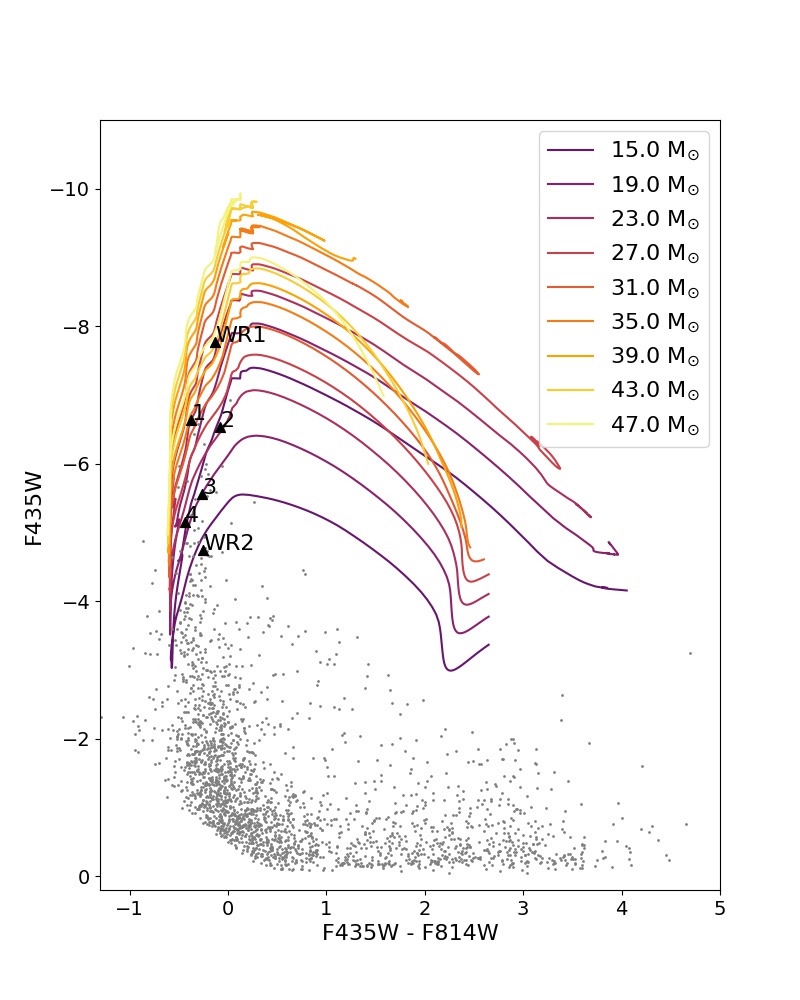}{0.52\textwidth}{(a)}
          \fig{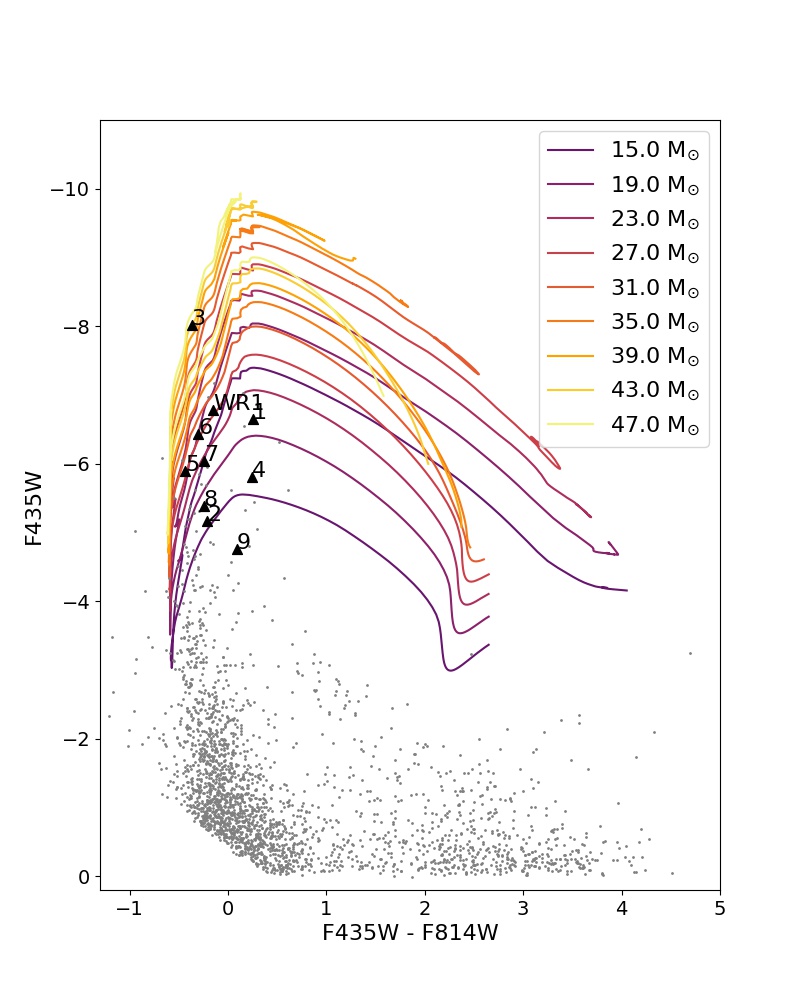}{0.52\textwidth}{(b)}
          }
\caption{Color-magnitude diagrams of D118 (left) and D119 (right) from the cropped ANGST star catalogs (within 22\arcsec of the cluster centers and with SNR $>$ 10, grey data points, see Section \ref{sec:hst}) with MIST evolutionary tracks \citep{choi16} (in the mass range $M = 15{-}47$~M$_\odot$) over-plotted on both panels. The O-type and WR stars in D118 and D119 are marked with black triangles. Unlabeled bright objects in the CMDs are B-type stars not analyzed in this paper. \label{fig:cmd}}
\end{figure*}

Fig.~\ref{fig:cmd} shows HST color-magnitude diagrams (CMDs) for $22\arcsec$ apertures around the central coordinates of the two HII region complexes, assuming a distance modulus of $26.5$ for NGC 300 (corresponding to a distance of 2 Mpc to NGC 300, \citealt{dalcanton09}). For completeness, in Fig.~\ref{fig:cmd} we show all stars, i.e.\ including those with $m_{\rm{F814W}}>22$, within 22\arcsec of the cluster centers ($2313$ and $2384$ stars for D118 and D119, respectively). Over-plotted are MIST\footnote{\hyperlink{http://waps.cfa.harvard.edu/MIST/}{MESA Isocrhones \& Stellar Tracks}} evolutionary tracks \citep{choi16,dotter16} in the mass range $M = 15{-}47$~M$_\odot$ (in steps of $4~{\rm M}_\odot$) for [Fe/H] = -0.5 \citep{richer85} and with A$_{\rm V} = 0.06$ \citep{burstein84}.

\section{Stellar spectroscopy and classification} \label{sec:stars}
At the distance of NGC 300, the seeing-limited spatial resolution of MUSE results in blending of single stellar sources in crowded regions, e.g.~star clusters (see Section \ref{sec:119_5_9}).   To extract stellar spectra from these regions and study their stellar content, we exploit the {\sc python} package {\sc PampelMuse} \citep{kamann13} to compute a wavelength-dependent PSF for the stellar sources in our MUSE cubes.  As described in \cite{zeidler18}, {\sc PampelMuse} performs PSF photometry of isolated bright sources in an IFU cube based on a high angular resolution photometric catalog.  If run to full completion, after the PSF fit (which also includes a WCS correction between the input catalog and the corresponding sources detected in the IFU data) {\sc PampelMuse} then propagates the PSF profile to all the stellar sources listed in the input catalog, estimates and subtracts the background around each star, and finally delivers a spectral catalog of all identified sources.

Given that the current version of {\sc PampelMuse} is not designed to perform background estimates for stars within HII regions, we are not able to use the output spectra, as the heavy contamination by nebular emission lines results in a faulty background subtraction. We therefore only use {\sc PampelMuse} to perform its first analysis stage, i.e.\ the PSF fit and WCS correction.  

We then perform a first spectroscopic analysis to identify the massive, feedback-driving stars\footnote{This study focuses on massive stars, as these dominate over intermediate- and low-mass stars in terms of feedback energetics.  We will discuss the extraction and classification of all stars down to the detection limit within these two regions in a forthcoming publication.} within the HII regions by evaluating the presence (or absence) of specific spectral lines characteristic of O-type and WR stars:

\begin{itemize}
    \item O-type stars: He II $\lambda$5411, He I $\lambda\lambda$4921, 5876, 6678, 7065, 7281;
    \item WR stars: broad He II $\lambda$5411 and He I $\lambda\lambda$4921, 5876, 6678, 7065, 7281 emission, C III $\lambda$5696 and C IV $\lambda$5808 (WC stars), N III $\lambda$8237 (WN stars), O V $\lambda$5835 (WO stars);
\end{itemize}

To evaluate the presence (or absence) of spectral lines, we crop the extracted spectra to $\pm50$~\AA\ around the central wavelength of each line, and perform a continuum fit and subtraction.  Ideally, the detection and significance of a line should be quantified by evaluating the presence of data points above a certain noise threshold. However, given that the stars of interest are embedded in HII regions and superimposed on a galactic background, their spectra are noisy.  To ensure that the above criteria do not lead to false negatives (and hence to an underestimate of the combined energy input of the massive stellar populations in the regions), we visually inspect each line to confirm its significance.  Moreover, we require the following conditions to be satisfied: (i) the width of the line has to be $>1.25$~\AA\  (i.e.\ greater than the spectral sampling of MUSE); (ii) we require the fitted line centroid to be within 2 times the spectral sampling (e.g.\ $5413.3\pm2.5$~\AA). Corresponding to a velocity of $\sim140$~km~s$^{-1}$, the latter assumes that the HII regions are young and that the stars are still associated with the surrounding gas, which is reasonable considering that the systemic velocities of D118 and D119 are ${\sim} 102$ and ${\sim} 105$~km~s$^{-1}$, respectively (derived from VLA HI data, \citealt{faesi14}). This also ensures that potential OB-runaways with velocities up to $\sim40$~km~s$^{-1}$ are not missed (e.g.\ \citealt{gies86}).

\begin{figure*}[ht!]
\epsscale{1.25}
\plotone{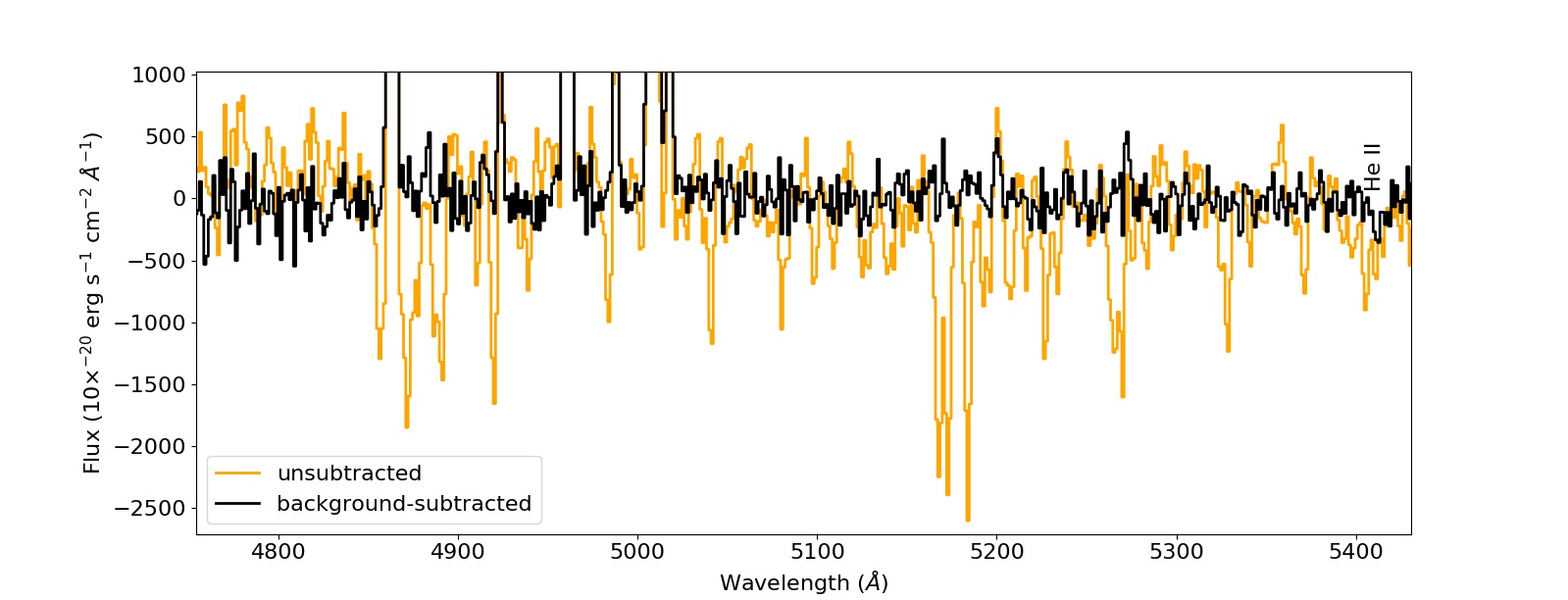}
\caption{Continuum-subtracted spectrum of the O-type star {[DCL89]}119-5 before (orange) and after (black) correcting for the unresolved galactic stellar background. The location of the He II $\lambda$5411 line, used to classify O stars, is indicated. Given the large wavelength range covered by MUSE ($\sim$ 4750 - 9350 \AA), we cropped the shown spectrum to a smaller (yet representative) portion to better illustrate the result of the galactic background subtraction. The level of noise in the continuum of the background spectrum is $\approx$ 60.}
\label{fig:bkg_spec}
\end{figure*}

The resulting stellar spectra are subsequently corrected for the unresolved galactic stellar background, which introduces absorption features of an old stellar population in the spectra of our young, massive stars of interest.  The galactic stellar background is estimated from the integrated spectrum of an aperture centered on a dark pixel (from the darkest $5\%$ of pixels from each cube, i.e.\ without nebular emission and not coinciding with stellar sources, as was used for the nebular emission analysis).  Because the stellar background is unresolved across the entire field-of-view, the dark region can be taken as representative for the background subtraction.  The background aperture is equal in size to the PSF-fitted aperture used to extract stellar spectra, hence we subtract a background spectrum integrated over the same number of pixels.  Fig.~\ref{fig:bkg_spec} shows the result of the stellar background correction of the O-type star [DCL89]119-5 (see Section~\ref{sec:119_5_9}): the uncorrected spectrum (orange) shows deep absorption features from the older galactic population along the line of sight, which lead to Balmer absorption and contamination of absorption lines relevant for spectral classification such as the He II $\lambda5411$ line.  

In \citetalias{mcleod18b} we use MUSE data to classify O-type stars in HII regions in the LMC based on the equivalent width (EW) of the helium lines (e.g.~\citealt{kerton99}).  This empirical method of determining the spectral type of massive, O-type stars relies on the EW ratio of the atomic and singly ionized He absorption lines, e.g. EW$_{\lambda4921}$/EW$_{\lambda5411}$.  Due to the stars being embedded in the ionized material of the HII regions, the He~I $\lambda4921$ line is unreliable for spectral typing, as it can be of nebular origin when in emission (e.g.\ \citealt{byler18}), therefore, contaminating the potential stellar absorption line.  Instead, we derive atmospheric parameters by fitting the He~II $\lambda5411$ absorption line with PoWR model atmospheres \citep{hainich19}. The PoWR grids come in Galactic, LMC and SMC metallicities for both WR and O-type stars. With $Z \approx 0.33~Z_{\odot}$ for NGC 300 \citep{butler04}, we use the PoWR LMC model atmospheres ($Z = 0.5~Z_{\odot}$), which are closest in metallicity to NCG 300 (see e.g. \citealt{ramcha18,ramcha19} for an application of PoWR models in the Magellanic Clouds).  We crop the modeled and the observed spectra of the identified O-type stars to a wavelength range of $5380{-}5440$~\AA\ and find the best fit model for each star via $\chi^{2}$ minimization,

\begin{equation}
    \chi^{2} = \sum_{i=1}^{N}\frac{(O_{i}-E_{i})^{2}}{\sigma_{i}}
\end{equation}

with {\it O} being the observed flux, {\it E} the model flux, and $\sigma$ the mean noise of the spectrum plus a random (Gaussian) noise component equivalent to the computed noise, and use a Monte-Carlo approach (for each star we perform 1000 runs) to estimate the 1$\sigma$ errors on the fitted parameters. For the ionizing photon fluxes associated with each grid model, as well as the WR luminositites (for which for these constant mass and luminosity is assumed throughout the PoWR grid models) we adopt a generous 20\% uncertainty.

In terms of stellar feedback, the main stellar parameter of interest for this work derived from the PoWR model atmosphere fit are the bolometric luminosities of the stars, as well as the amount of ionizing (Lyman continuum) photons emitted per second, $Q_{H}$ (Columns 4 and 7 in Table \ref{tab:ostars}). In this respect, there are three major sources of uncertainty that we briefly discuss here. First, we note wind contamination of the He~II$\lambda$5411 line: this is mainly problematic in the case of extreme supergiants, which we do not observe in the two regions.  Second, we do not account for unresolved binaries, and the derived photon fluxes are therefore lower limits.  Third, in a $\sim20$~\AA\ proximity of the He~II$\lambda$5411 line, there are several metal lines from the galactic background, the subtraction of which is a potential source of uncertainty (i.e.~by decreasing the amplitude of the He~II line).  In the background subtraction illustrated in Fig.~\ref{fig:bkg_spec}, the S/N ratio of He II line at 5411 \AA\ improves from $\sim$1.1 prior to $\sim$2.2 after the background correction, showing that over-subtraction is not problematic. We furthermore note that the He~II$\lambda$5411 absorption line is not surrounded by (or blended with) any nebular emission line, hence we do not expect the nebulosity to affect the stellar classification. 

However, the main problem in deriving the stellar parameters is the uniqueness of fitting: for an observed depth of the He~II absorption line there is a suite of models from the grid with different sets of surface gravity, mass, and temperature that fit (see Fig.~A\ref{fig:powr}). In addition to the $\chi^{2}$ minimization, in the next sections we therefore evaluate the validity of obtaining stellar parameters from the PoWR model fits by comparing the effective temperatures and luminosities (see Table \ref{tab:ostars}) derived from the PoWR models with those obtained (by eye) from the HST CMD (see Fig.~\ref{fig:cmd}).  The combination of the MUSE spectroscopy and the HST photometry therefore allows us to reliably derive stellar parameters.  In the feedback analysis presented in Section \ref{sec:gas}, we proceed in combining photon flux and luminosity of the single stars to summed fluxes and luminosities in each region (depending on their stellar content), and we will discuss the uncertainty on the combined parameters flux at that stage.   

Spectral contamination by nebular emission lines becomes an issue when fitting PoWR models to the WR spectra, as the classification of these rely on the (broad) emission features characteristic for these kind of stars. The nebular contamination is in the form of narrow emission lines superimposed on the broad WR lines, and we remove these prior to classification by fitting two-component Gaussians to the emission features and subsequently removing the narrow (nebular) component.

\subsection{Stars in the D118 and D119 complexes}\label{sec:regionstars}
In the two complexes analyzed, the classification results in the identification of $13$ O-type stars (4 in the D118 complex and 9 in D119), and $4$ WR stars (2 in D118 and 2 in D119). Two of the WR stars have been previously identified \citep{breysacher97,schild03}, validating our classification process. Of the $13$ O-type stars, none were previously known (e.g.\ \citealt{bresolin02}), making these first detections and classifications. The coordinates and best-fit PoWR model parameters are listed in Tables~\ref{tab:ostars} and \ref{tab:wrstars}. The last column in Table \ref{tab:ostars} also lists the approximate ($\pm$3 M$_{\odot}$) mass for each star as derived from the MIST evolutionary tracks shown in Fig.~\ref{fig:cmd}.  

\begin{figure*}[h]
\gridline{\fig{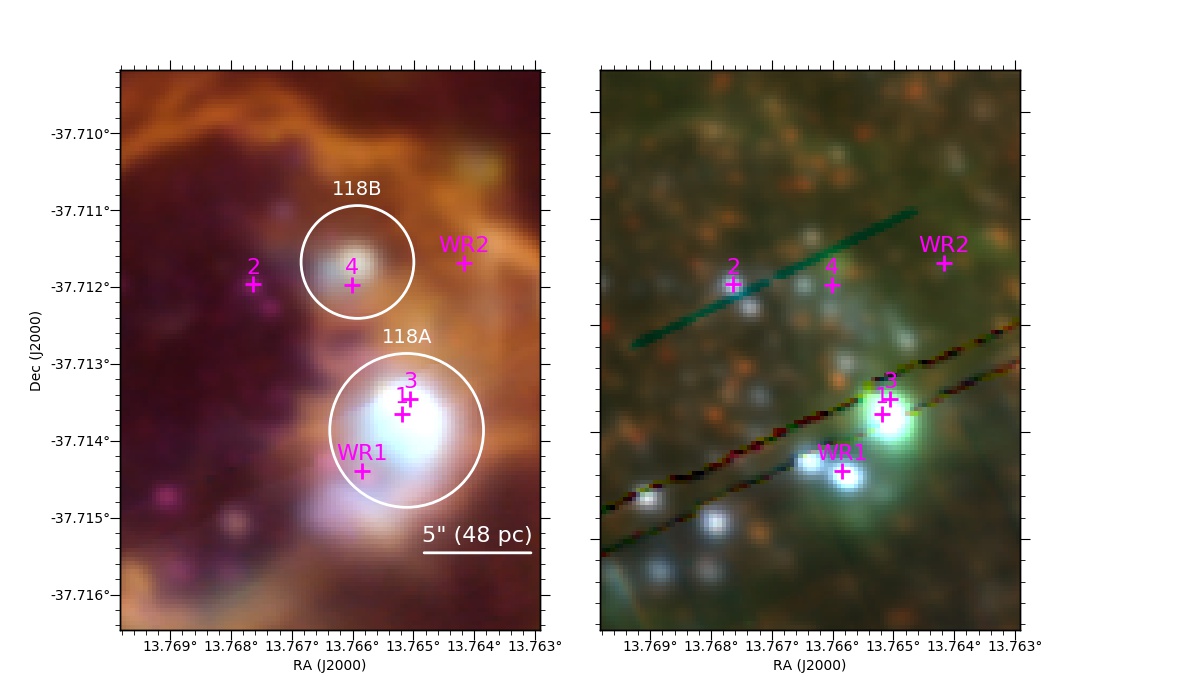}{0.95\textwidth}{(a)}}
\gridline{\fig{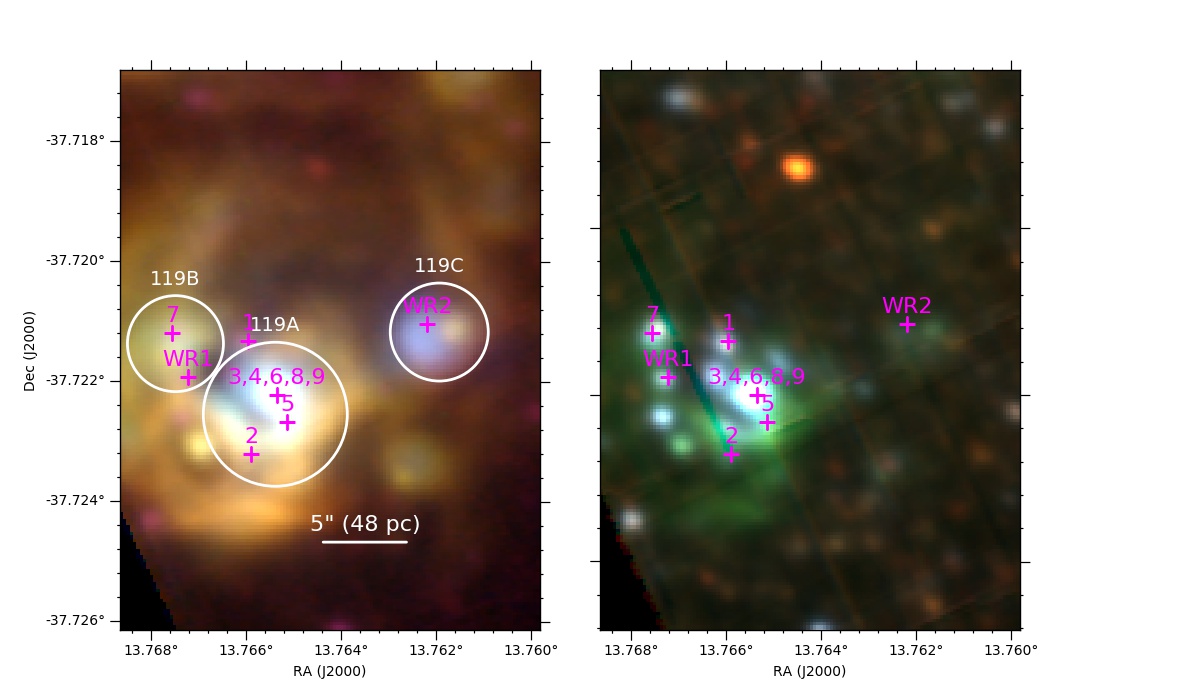}{0.95\textwidth}{(b)}}
\caption{[DCL88]118 (a) and [DCL88]119 (b) zoom-ins showing the location of the identified O-type and WR stars (magenta crosses).  The false-color composites in both panels are the same as in Fig.~\ref{rgb} (on the left, red: [SII]$\lambda$6717; green: H$\alpha$; blue: [OIII]$\lambda$5007, on the right, arbitrary $RGB$). White circles indicated the five individual HII regions within the D118 and D119 complexes.  In panel (b) sources 119-3, 4, 5, 6, 8, and 9 are marked by a single cross, a zoomed-in version of this image is shown in Fig.~\ref{blending}. Image artefacts (e.g. upper right panel) are due to mosaicing and do not affect the analysis.}
\label{fig:ostars}
\end{figure*}

\begin{figure}[ht!]
\epsscale{1.3}
\plotone{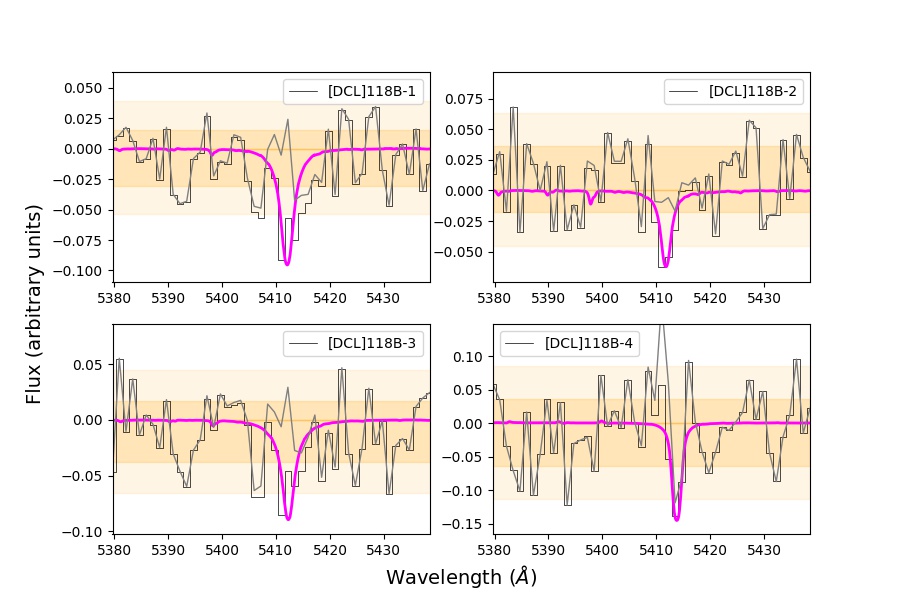}
\caption{He II $\lambda$5411 line of the four D118 O-type stars (black), the best-fit PoWR model atmosphere (magenta), and residulas (gray).  See Section~\ref{sec:stars} and Table~\ref{tab:ostars} for details.}
\label{fig:118Oatmo}
\end{figure}

Fig.~\ref{fig:ostars} shows the spatial distribution of the identified O and WR stars relative to the HII region complexes, as shown by ionized gas emission (left) and stellar RGB (right). The top row of Fig.~\ref{fig:ostars} shows the D118 complex, which includes two HII regions (118A, 118B; highlighted with white circles). The ionized gas emission reveals a large ($\sim$140~pc) bubble from the center of which no gaseous emission is detected, a thin rim mainly seen in [SII] emission (see Fig.~\ref{rgb}), and a higher-ionized region to its south-west. It is in the center of this highly-ionized region that the O and WR stars are found, mostly spatially coincident with the individual HII regions 118A and 118B.  The spectra of the identified massive stars (cropped to the relevant He II absorption line) are shown in Fig.~\ref{fig:118Oatmo}, together with the best fit PoWR models (magenta). We note that there are at least four candidate O-type stars in D118, but no reliable fit could be obtained, because He II$\lambda$5411 emission is superimposed on the stellar absorption.  We therefore do not include these in the analysis.

\begin{figure}[ht!]
\epsscale{1.3}
\plotone{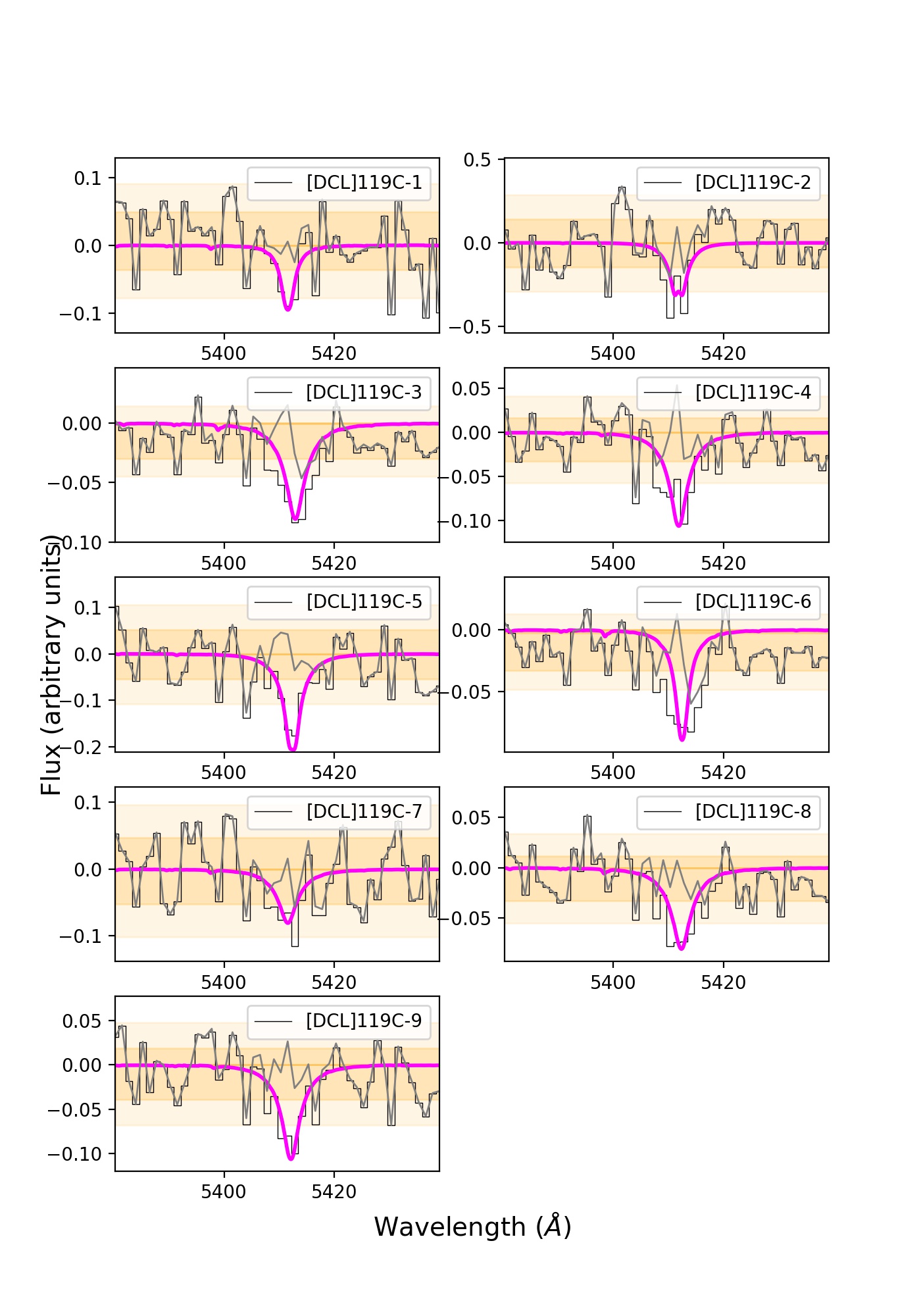}
\caption{He II $\lambda5411$ line of the nine O-type stars in D119 (black), the best-fit PoWR model atmosphere (magenta), and residuals (gray).  See Section~\ref{sec:stars} and Table~\ref{tab:ostars} for details.}
\label{fig:119Oatmo}
\end{figure}

The bottom panels of Fig.~\ref{fig:ostars} show the three HII regions in the D119 complex (i.e.~119A; 119B; 119C; white circles).  As for D118, no O-type stars are found in the literature for this HII region complex, while there are two WR stars, one confirmed and one candidate (sources \#38 and \#33 in \citealt{schild03}, respectively).  From the presence of the He II$\lambda$5411 absorption line we identify five O-type stars within D119 (see Table \ref{tab:ostars}).  The spectra of the D119 O-type stars are shown in Figure \ref{fig:119Oatmo}, together with the best fit PoWR model atmospheres (magenta).  We find two candidate O-type stars that show He I in absorption, but either have contaminated He~II absorption lines or lack reliable ones altogether, and are therefore not included in the analysis of this paper.

In the following, we discuss the classification of the individual sources in each region.

\begin{deluxetable*}{lcCccccc|c}
\tablecaption{Model stellar parameters for the O-type stars reported from the best-fit PoWR atmospheres \citep{hainich19} (with Column 3 indicating the best-fit model name). The last column lists approximate stellar masses as derived (by eye) from the CMD.  The fitted spectra are shown in Figures~\ref{fig:118Oatmo} and \ref{fig:119Oatmo}. Uncertainties on the fitted parameters correspond to the 1$\sigma$ standard deviation from 1000 Monte-Carlo runs.\label{tab:ostars}}
\tablecolumns{9}
\tablenum{3}
\tablewidth{0pt}
\tablehead{
\colhead{ID} & \colhead{Coordinates} & \colhead{PoWR model}  & \colhead{log $L$} & \colhead{log $T_{\rm{eff}}$} & \colhead{log $g$} & \colhead{log $Q_{\rm H}$} & \colhead{$M$}& \colhead{$M_{\rm MIST}$}\\
\colhead{} &  \colhead{(J2000)} & \colhead{} &  \colhead{(L$_{\odot}$)} & \colhead{(K)} & \colhead{[cgs]} & \colhead{(s$^{-1}$)} & \colhead{(M$_{\odot}$)} & \colhead{(M$_{\odot}$)}}
\startdata
{[DCL88]}118-1 & 0:55:03.63 -37:42:49.05 & 30-36 & 5.0$\pm$4.4 & 4.48$\pm$3.57 & 3.6$\pm$0.3 & 48.08$\pm$47.61 & 21.5$\pm$0.8 & 23 \\
{[DCL88]}118-2 & 0:55:04.24 -37:42:43.06 & 28-34 & 5.1$\pm$4.2 & 4.45$\pm$3.60 & 3.4$\pm$0.4 & 47.98$\pm$47.58 & 23.0$\pm$0.7 & 24 \\
{[DCL88]}118-3 & 0:55:03.61 -37:42:48.46 & 31-38 & 4.8$\pm$4.3 & 4.49$\pm$3.67 & 3.6$\pm$0.3 & 47.92$\pm$47.45 & 18.8$\pm$1.2 & 19 \\
{[DCL88]}118-4 & 0:55:03.84 -37:42:43.12 & 28-30 & 5.9$\pm$5.3 & 4.47$\pm$3.35 & 3.0$\pm$0.2 & 49.15$\pm$48.44 & 49.2$\pm$9.6 & 17 \\
\hline
{[DCL88]}119-1 & 0:55:03.85 -37:43:16.68 & 30-36 & 5.0$\pm$4.9 & 4.48$\pm$3.36 & 3.6$\pm$0.3 & 48.08$\pm$47.38 & 21.5$\pm$8.1 & 20 \\
{[DCL88]}119-2 & 0:55:03.84 -37:43:23.43 & 44-40 & 5.6$\pm$4.8 & 4.64$\pm$3.47 & 4.0$\pm$0.2 & 49.41$\pm$48.60 & 47.5$\pm$4.1  & 16 \\
{[DCL88]}119-3 & 0:55:03.73 -37:43:19.93 & 33-44 & 4.3$\pm$4.2 & 4.52$\pm$3.28 & 4.4$\pm$0.3 & 47.38$\pm$47.06 & 15.7$\pm$4.6 & 43 \\
{[DCL88]}119-3\tablenotemark{a} & ... & 33-34 & 5.7$\pm$5.3 & 4.52$\pm$3.82 & 3.4$\pm$0.7 & 49.21$\pm$48.51 & 41.3$\pm$8.3 & 43 \\
{[DCL88]}119-4 & 0:55:03.70 -37:43:19.42 & 33-42 & 4.5$\pm$3.9 & 4.52$\pm$3.28 & 4.2$\pm$0.3 & 47.70$\pm$47.25 & 17.2$\pm$5.4 & 16 \\
{[DCL88]}119-5 & 0:55:03.67 -37:43:21.53 & 36-42 & 4.8$\pm$4.3 & 4.56$\pm$3.50 & 4.2$\pm$0.3 & 48.21$\pm$47.67 & 21.6$\pm$8.6 & 19 \\
{[DCL88]}119-6 & 0:55:03.73 -37:43:19.79 & 31-38 & 4.8$\pm$4.4 & 4.49$\pm$3.03 & 3.8$\pm$0.3 & 47.92$\pm$47.29 & 18.8$\pm$3.9 & 19\\
{[DCL88]}119-7 & 0:55:04.28 -37:43:15.83 & 33-44 & 4.3$\pm$4.2 & 4.52$\pm$3.29 & 4.4$\pm$0.4 & 47.38$\pm$46.56 & 15.7$\pm$6.8 & 15\\
{[DCL88]}119-8 & 0:55:03.78 -37:43:18.91 & 33-44 & 4.3$\pm$3.7 & 4.52$\pm$3.29 & 4.4$\pm$0.3 & 47.38$\pm$46.00 & 15.7$\pm$4.8 & 16\\
{[DCL88]}119-9 & 0:55:03.77 -37:43:18.73 & 33-42 & 4.5$\pm$3.5 & 4.52$\pm$3.23 & 4.2$\pm$0.3 & 47.70$\pm$47.96 & 17.2$\pm$5.3 & $<15$\\
\enddata
\tablenotetext{a}{Revised parameters of 119-3 obtained by additionally fitting the He II $\lambda6406$ line, see Section~\ref{sec:119_34}.}
\end{deluxetable*}

\subsubsection{[DCL88]118-1}\label{sec:118_1}
With an absolute magnitude of M$_{F435W}$ = -6.63, this source is among the brightest in 118A, and it is associated with bright [OIII]$\lambda$5007 emission (see Fig.~\ref{fig:ostars}).  On the CMD it is identified as either a young, massive ($\sim$ 33 M$_{\odot}$, log(L/L$_{\odot}$) $\approx$ 5.2) star or as an older ($\sim$ 8 Myr) and less massive ($\sim$ 23 M$_{\odot}$, log(L/L$_{\odot}$) $\approx$ 5.3) star.  PoWR model fitting yields a mass of 21.5 M$_{\odot}$ and a luminosity of log(L/L$_{\odot}) \sim$ 5.0, hence favoring the less massive MIST track, and also consistent with an age of $\sim$ 4 Myr reported by \citet{faesi14}.

\subsubsection{[DCL88]118-2}\label{sec:118_2}
Located further inside the main bubble of the D118 complex than the other O-type stars identified in this region, the model atmosphere fit of 118-2 is consistent with the stellar parameters derived from the CMD, i.e.~consistent with a luminous $\sim$ 23 M$_{\odot}$ star.

\subsubsection{[DCL88]118-3}\label{sec:118_3}
In the MUSE data, 118-3 and 118-1 are blended into one luminous source in the 118A region on the south-western rim of the giant D118 bubble.  Given the high spatial resolution HST input catalog, we are able to de-blend the two stars (with a separation of approximately 20 times the spatial resolution of HST/ACS, and F435W and F814W signal-to-noise ratios of 249.8 and 332.9, respectively, these are therefore clearly two distinct stars) and obtain separate stellar parameters.  We find, in agreement with the HST photometry, that 118-3 is a $\sim$ 18.8 M$_{\odot}$ star with a luminosity of log(L/L$_{\odot}$) $\sim$ 4.9.

\subsubsection{[DCL88]118-4}\label{sec:118_4}
From the model atmosphere fitting routine, this O-type star appears to be the most massive source in the D118 complex.  However, this is not reflected in the CMD, where 118-4 appears to be less luminous than the other three O-type stars in the region, and is more consistent with MIST tracks in the 19 -- 23 M$_{\odot}$ range, rather than the 49.2 M$_{\odot}$ mass derived from the model fit.  Given an extinction of A$_{V} \sim$ 0.5 (from the MUSE H$\beta$ and H$\alpha$ lines) in that region and the fact that this star has formed its own small bubble of is highly-ionized gas in the form of [OIII]$\lambda$5007 emission (see upper left panel in Fig.~\ref{fig:ostars}), we suggest that this is indeed a high-mass source as obtained from the PoWR model fitting, and therefore adopt the derived photon flux for the feedback analysis discussed in Section \ref{sec:gas}.

\subsubsection{[DLC88]118-WR1 and [DCL88]118-WR2}\label{sec:118_wr}
\begin{figure*}[ht!]
\gridline{\fig{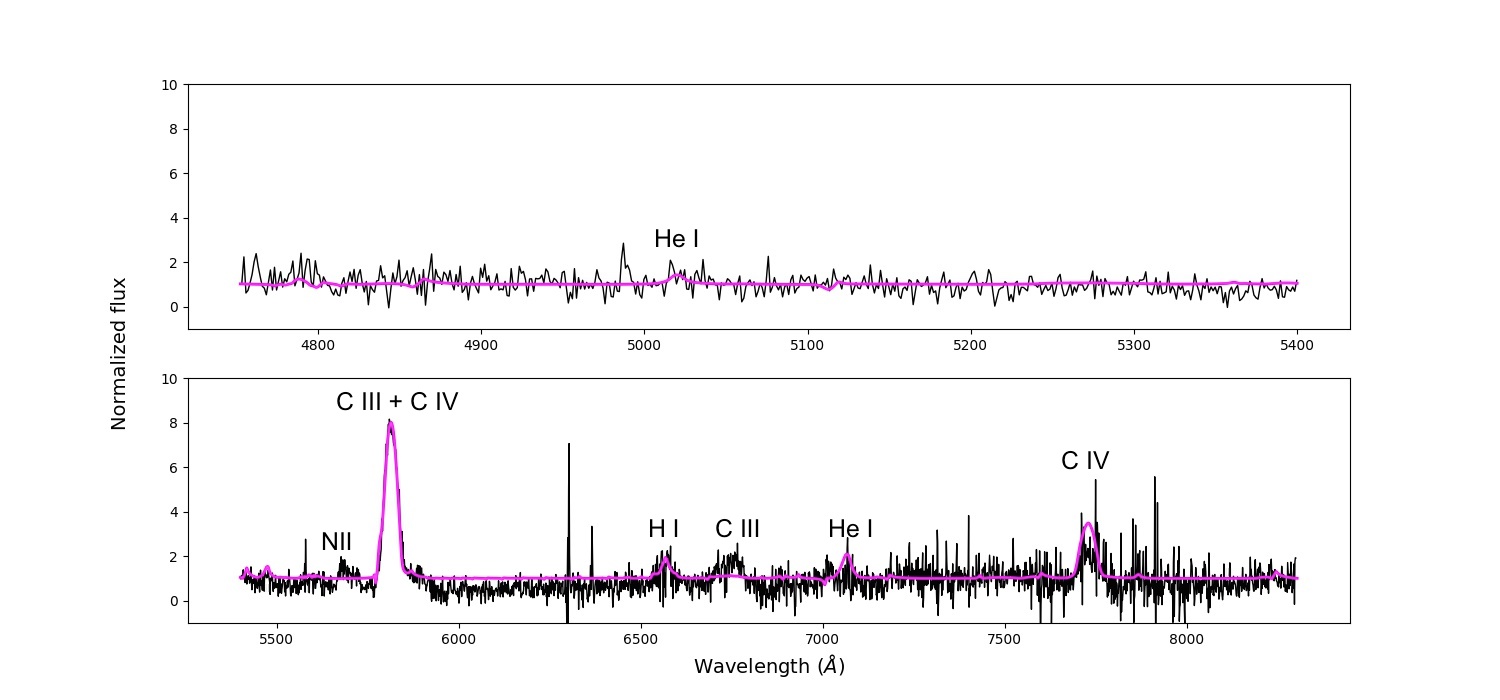}{0.95\textwidth}{(a)}}
\gridline{\fig{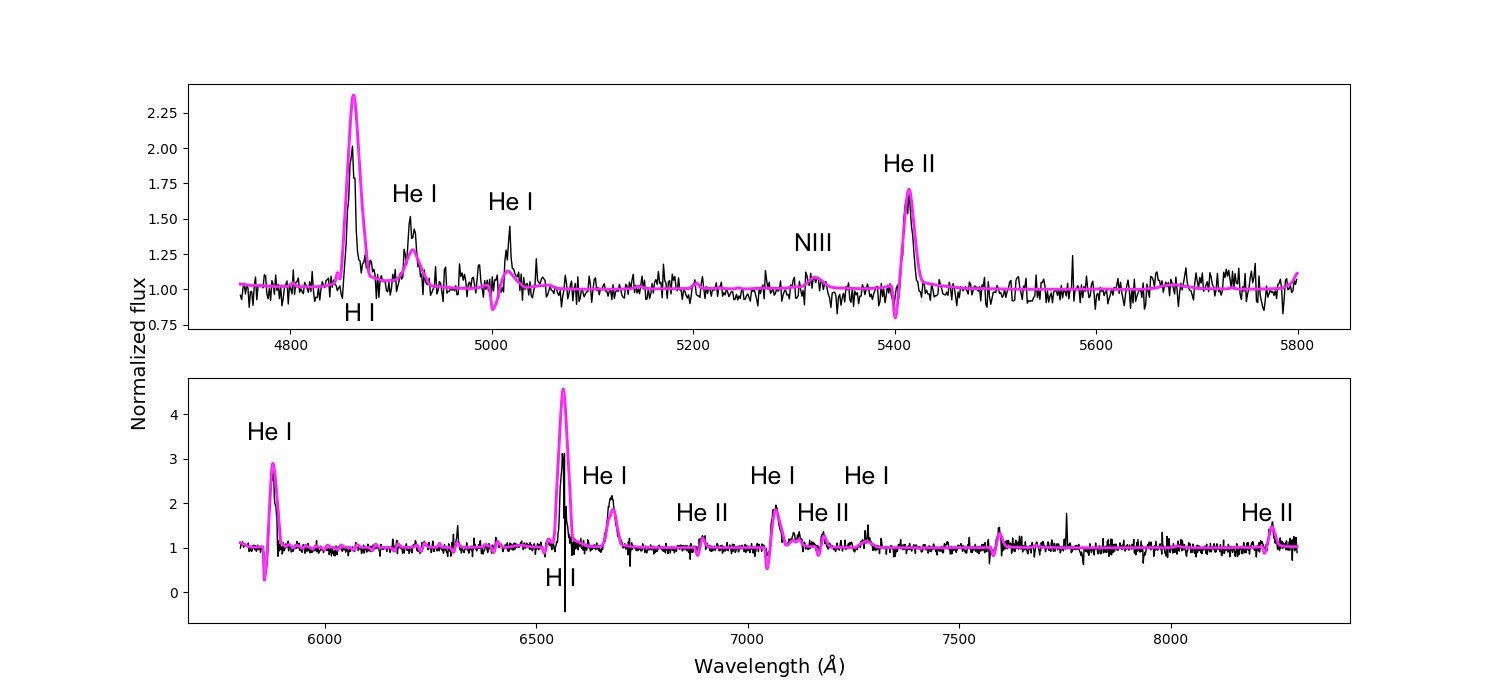}{0.95\textwidth}{(b)}}
\caption{Spectra (black) of the WR stars 118-WR2 (a) and 119-WR1 (b), and the best-fit PoWR model atmospheres (magenta).  See Section~\ref{sec:stars} and Table~\ref{tab:wrstars} for details. Prominent lines are labeled.}
\label{fig:wr119}
\end{figure*}

Both of the D118 WR stars, 118-WR1 and 118-WR2, are listed in \citet{schild03} (their sources \#37 and \#34, respectively), and while 118-WR2 is classified as a WC5-6 star in that study, 118-WR1 is listed as a candidate WR star.  MUSE spectra for these two sources are shown in Figures \ref{fig:wr119} (panel a) and \ref{fig:wr_unclass} (black spectrum), respectively.  We fit the spectrum of 118-WR2 to the WC PoWR grid and obtain a temperature of T $\sim$ 79.4 kK and a luminosity of $\log({\rm L}/{\rm L}_{\odot}) = 5.3$, which are more consistent with a WC7 spectral type than with a WC5-6 type \citep{crowther07}.

118-WR1 shows a dominant C IV feature at $\lambda$5800 \AA, and aside from some possible Fe III absorption in the 5900 - 6000 \AA\ range, there are no other WR emission features.  Like 119-WR2 (see Section \ref{sec:119_wr}), due to the absence of O V $\lambda$5590, C III $\lambda$5696, and He II $\lambda5411$ emission, we suggest that this star is an early (e.g.~WC4) WC star \citep{gray09}, and therefore assume an ionizing photon flux of log Q$_{H}$ = 49.50 and a luminosity $\log(\rm{L}/{\rm L}_\odot) = 5.6$ s$^{-1}$ \citep{crowther07,crowther02} for our feedback analyses.

\subsubsection{[DCL88]119-1}\label{sec:119_1}
Compared to the other sources, 119-1 suffers the least from nebular contamination. From the best fit PoWR model we find a mass of 21.5 M$_{\odot}$, which is in good agreement with a tentative mass of 20 M$_{\odot}$ obtained from the HST CMD. 

\subsubsection{[DCL88]119-2}\label{sec:119_2}
The spectrum of 119-2 shows emission superimposed on the He II absorption line, indicating possible nebular contamination.  Indeed, the best-fit PoWR model yields a stellar mass that is almost three times what is derived from the MIST evolutionary tracks. 

\subsubsection{[DCL88]119-3 and [DCL88]119-4}\label{sec:119_34}
\begin{figure}[ht!]
\plotone{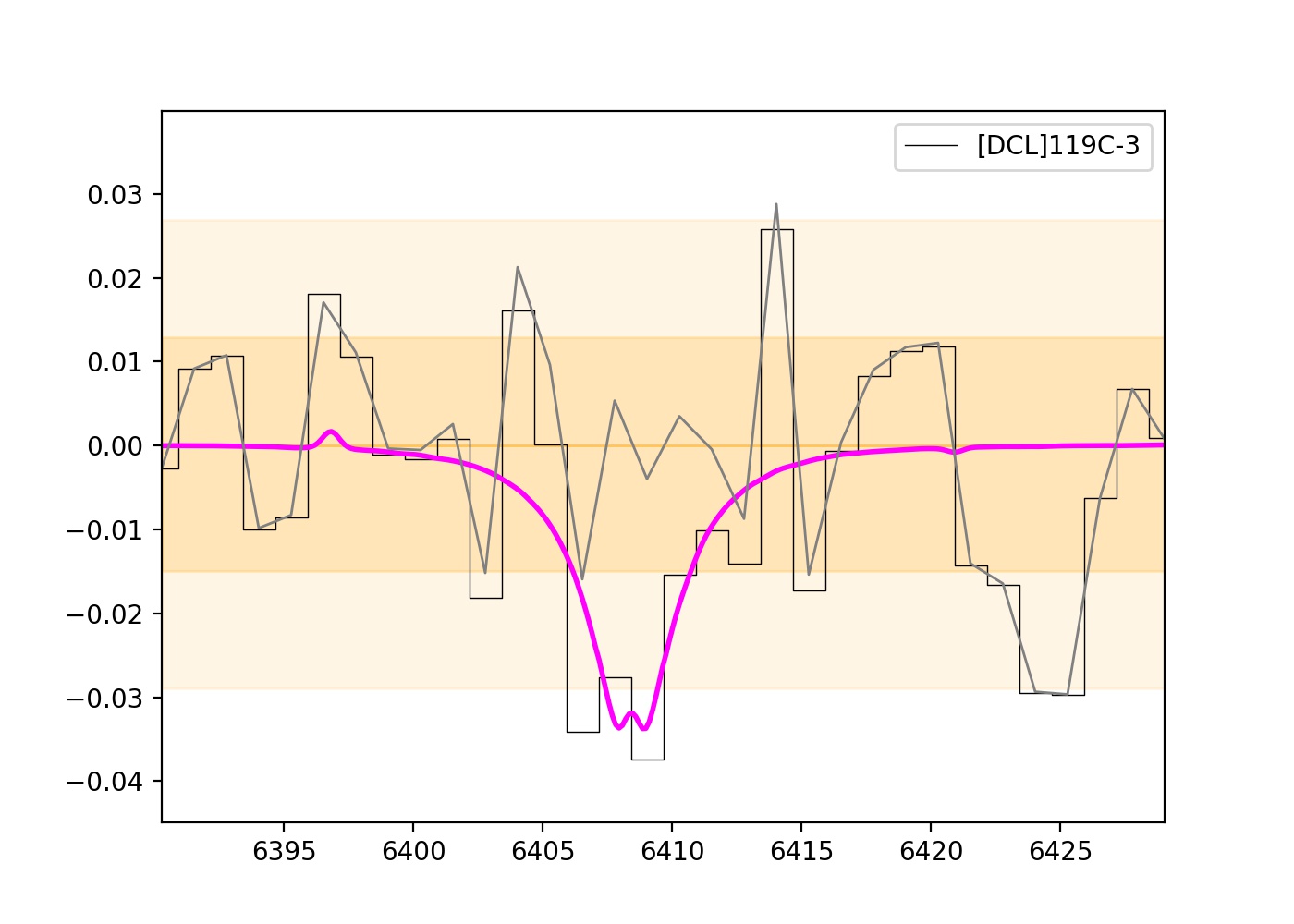}
\caption{He II $\lambda$6406 line of the O-type star [DCL88]119-3, the best-fit PoWR model (magenta), and the residuals (gray). See Section \ref{sec:119_34}.}
\label{fig:119c3}
\end{figure}

These two stars are located in the region with the highest degree of ionization, traced by strong [OIII] $\lambda5007$ emission, and are most likely the main ionizing sources of the HII region.  The PoWR mass derived for 119-4 is in good agreement with the MIST mass as per the CMD.  However, for 119-3, the mass derived from the model atmosphere fit is clearly underestimated, given that the derived parameters would place 119-3 approximately 5 mag fainter on the CMD.  To further investigate the stellar parameters of 119-3, we evaluate the presence of He II absorption lines redder than the $\lambda5411$ line\footnote{[DCL88]119-3 is the only O star in our sample showing He II lines other than the $\lambda$5411 line.}, as these become prominent for masses $\gtrsim15$~M$_{\odot}$. Of these, only the He II $\lambda6406$ line is isolated enough from nebular emission lines to guarantee a good fit. The best-fit PoWR model to the He II $\lambda6406$ line (shown in Fig.~\ref{fig:119c3} and listed in Table~\ref{tab:ostars}) yields a mass of $M = 41.3~{\rm M}_{\odot}$ and a luminosity of $\log({\rm L}/{\rm L}_{\odot}) = 5.2$, which are in good agreement with the HST photometry.

\subsubsection{[DCL88]119-5 to [DCL88]119-9 }\label{sec:119_5_9}
\begin{figure*}[ht!]
\plotone{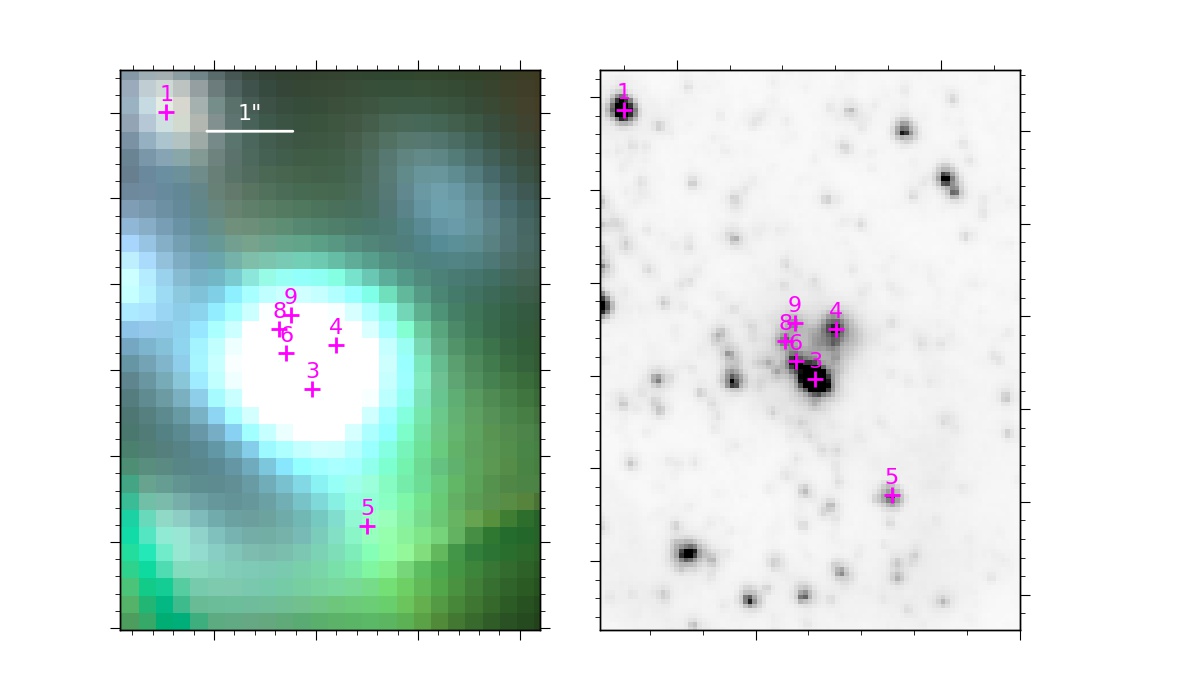}
\caption{Zoom-in of the centre of 119A illustrating the effect of lower spatial resolution of the MUSE data. The left panel is a MUSE $\mathit{VRI}$ false-color image (see Section~\ref{sec:musedata}, resolution $\sim$ 1\farcs0), while the right panel shows the corresponding HST {\it F814W} image ($\sim$ 0\farcs13/pixel spatial resolution) from the ANGST survey \citep{dalcanton09}. Magenta crosses correspond to O-type stars identified in this work (see Section~\ref{sec:stars}).}
\label{blending}
\end{figure*}

These O-type stars are all in the lower mass range (15~M$_{\odot} < M < 20$~M$_{\odot}$), and the PoWR model masses and luminosties are in agreement with the stars' location on the CMD.  While 119-7 is located in the 119B HII region, 119-6, 8 and 9 are located in 119A, together with the stars 119-3 and 119-4, in what appears to be the brightest region of the D119 complex.  This crowded region, better visible in Fig.~\ref{blending}, is of particular importance for this work: the fact that we can de-blend the stars in this region despite the insufficient spatial resolution of the MUSE data by combining it with the high angular resolution HST catalog, demonstrates the true power of this method in extracting and classifying spectra at distances of a few Mpc from ground-based IFU data. Moreover, it also allows for robust spectroscopic modeling of WR, which are generally identified photometrically at these distances (see Section \ref{sec:findwr}).

\subsubsection{[DCL88]119-WR1 and [DCL88]119-WR2}\label{sec:119_wr}
With the MUSE data we confirm two WR stars in D119.  119-WR1, previously identified \citep{schild92,breysacher97,schild03} and classified as a WN9-10 star in \citet{schild03} (source \#38 in their table~1) shows strong H, He I and He II emission, as well as [NIII].  The best-fit H-rich PoWR model yields a luminosity of $\log(\rm{L}/{\rm L}_\odot) = 5.3$, $T = 44{,}700$~K and a mass loss rate of $-4.79$~M$_{\odot}$~yr$^{-1}$, consistent with a WN8 star with a massive ($>65$~M$_{\odot}$) progenitor \citep{crowther07}. 

\begin{figure}[ht!]
\epsscale{1.2}
\plotone{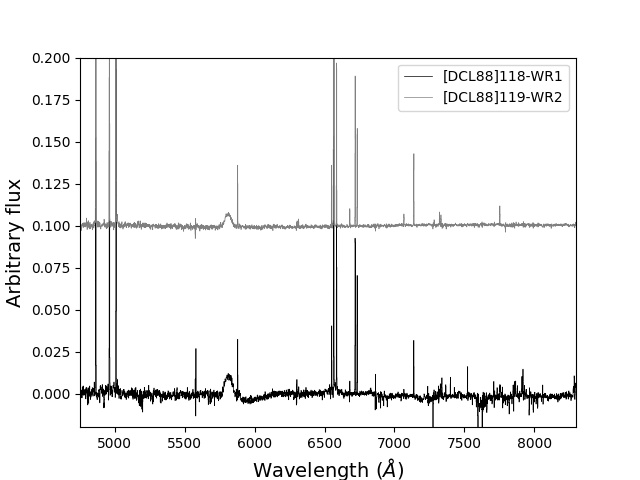}
\caption{Spectra of the WC stars 118-WR1 (bottom black) and 119-WR2 (top gray). See Section~\ref{sec:stars} and Table~\ref{tab:wrstars} for details.\label{fig:wr_unclass}}
\end{figure}

119-WR2, a WR candidate in \citet{schild03} (their source \#33), is located ${\sim}0.\arcmin2$ (${\sim}116$~pc) north-west of the approximate center of the main bubble, and is associated with highly ionized ([OIII]) gas, see Fig.~\ref{fig:ostars}. Except for the C IV bump at ${\sim}5800$~\AA, the spectrum of the WC star 119C-WR2 (see Fig.~\ref{fig:wr_unclass}) does not show any other strong WC feature, e.g. O V $\lambda$5590, C III $\lambda$5696, and He II $\lambda$5411, which would indicate WC 5 and later stars \citep{gray09}.  We suggest that this star is therefore likely an early (e.g.~WC4) WC star and assume an ionizing photon flux of $\log\rm{Q}_{\rm H}/{\rm s}^{-1} = 49.50$ and a luminosity $\log(\rm{L}/{\rm L}_\odot) = 5.6$ \citep{crowther07,crowther02}for our feedback analyses. We note however that for 119-WR2, as well as for 118-WR1, bluer spectroscopy would be required as to cover the strong C III, C IV bump at 4650 \AA\, which would allow a PoWR model fit.

\begin{deluxetable*}{lcCcclcr}
\tablecaption{Wolf-Rayet stars. \label{tab:wrstars}}
\tablecolumns{8}
\tablenum{4}
\tablewidth{0pt}
\tablehead{
\colhead{ID} & \colhead{Coordinates} & \colhead{PoWR model\tablenotemark{a}}  & \colhead{log $L$} & \colhead{log $T_{\rm{eff}}$} & \colhead{log $Q_{\rm H}$} & \colhead{$\dot{M}$} & \colhead{Type}\\
\colhead{} &  \colhead{(J2000)} & \colhead{} &  \colhead{(L$_{\odot}$)} & \colhead{(K)} & \colhead{(s$^{-1}$)} & \colhead{(M$_{\odot}$ yr$^{-1}$)} &  \colhead{}
}
\startdata
{[DCL88]}118-WR1 & 0:55:03.78 -37:42:51.80 & - & 5.6\tablenotemark{b} & - & 49.50\tablenotemark{b} & - & early WC\\
{[DCL88]}118-WR2 & 0:55:03.41 -37:42:42.14 & 09-10 & 5.3 & 4.90 & 49.18 & -5.39 & WC7\\
\hline
{[DCL88]}119-WR1 & 0:55:04.17 -37:43:18.91 & 06-13 & 5.3 & 4.65  & 49.10 & -4.79 & WN8 \\
{[DCL88]}119-WR2 & 0:55:02.94 -37:43:15.96 & - & 5.6\tablenotemark{b} & - & 49.50\tablenotemark{b} & - & early WC \\
\enddata
\tablenotetext{a}{\citet{hainich19}}
\tablenotetext{b}{Sources 118-WR1 and 119-WR2 do not have best-fit PoWR models, the ionizing photon flux is an assumption (see text Section \ref{sec:stars}).}
\end{deluxetable*}

\section{Integrated properties and stellar feedback} \label{sec:gas}
With knowledge of the massive stellar content of the 5 HII regions in D118 and D119, we now proceed to link the feedback-driving stars to the feedback-driven gas within the regions.  As mentioned in Section~\ref{sec:intro}, significant progress in the field of stellar feedback can only be obtained if a large enough sample of HII regions and their stellar contents is available, such that feedback can be analyzed over a meaningful dynamic range of HII region properties and massive stars.  Based on the 5 HII regions discussed here, in this work we develop the necessary analyses tools and methods, which can then be applied to large IFU data sets (see Section~\ref{sec:outlook}). 

Our analysis follows \citetalias{mcleod18b}, where we derived a detailed census of the massive stellar populations and feedback-related quantities for a total of $11$ individual HII regions in the LMC.  We then compare feedback properties derived with the same methodology across the LMC and NGC 300 environments that differ in terms of metallicity, galactic host, and stellar population.

For example, in \citetalias{mcleod18b} we find that by translating the spectral type of the O stars to an ionizing photon flux, $Q_{0,*}$, and relating this to the observed H$\alpha$ luminosity, $L_{\rm H\alpha}$, of each region, it appears that the massive stars in the LMC are less efficient at producing the observed $L_{\rm H\alpha}$ than prescribed by the typical Case~B recombination relation generally assumed for Galactic and extragalactic observations.  In \citetalias{mcleod18b} we suggest that a possible explanation for this discrepancy between $Q_{0,*}$ and $Q_{\rm H\alpha}$ is a consequence of {\it photon leakage}, i.e.\ a measurable fraction of ionizing photons emitted by the massive stars escape from the regions, and instead of contributing to the heating and ionization budget of the HII regions, they escape and contribute to the overall energetics of the host galaxy. As a result, the `feedback efficiency' describing the fraction of energy and momentum imparted on the surrounding interstellar medium (ISM) is considerably smaller than unity. However, from the LMC study alone it is not clear whether this is possibly related to the low-metallicity environment of the LMC, motivating a comparison of regions with different metallicities (e.g.\ MW, LMC, SMC, NGC 300) as performed here.

\subsection{Physical HII region parameters}\label{sec:params}
As in \citetalias{mcleod18b}, we derive HII region radii from the radial profiles of the continuum-subtracted MUSE H$\alpha$ maps, so that $R_{90}$ is the radius that encompasses $90\%$ of the H$\alpha$ flux (see Appendix \ref{sec:r90}). Given the different morphological appearances of the various HII regions, this phenomenologically-defined radius ensures that uncertainties between regions are mitigated (e.g.,~\citealt{lopez14}). We use the derived radii to extract the corresponding integrated spectrum for each region, and measure emission line intensities from Gaussian fits to the main nebular lines.  The emission line fluxes are corrected for (intrinsic) extinction with the reddening correction routine of {\sc PyNeb} \citep{luridiana15}, assuming $R_{V} = 3.1$ and a Galactic extinction curve \citep{cardelli89}. 
The extinction-corrected line fluxes are then used to derive electron densities and temperatures (from the [NII] and [SII] lines, respectively, as in \citealt{m16}). All physical parameters described in this section are listed in Table~\ref{tab:strucprop}.  

Overall, we find electron temperatures of the order of $10^{4}$~K, as well as low electron densities, i.e.~$n_{\rm{e}}<100$~cm$^{-3}$.  We compare the quantities derived for 118A to values from \citet{toribio16} who use optical spectroscopy to derive C and O abundances for seven regions in NGC 300.  With their region R23 corresponding to a small ($5.4\arcsec{\times}3\arcsec$) region around the O stars 118-1 and 118-3, we find good agreement between their electron density and temperature and those derived in this work. Specifically, these are $n_{\rm{e}} = 70\pm8$~cm$^{-3}$ and $T_{\rm{e}} = 9586\pm257$ K vs.~$n_{\rm{e}}<100$ cm$^{-3}$ and $T_{\rm{e}} = 8570\pm1300$ K for this work and \citeauthor{toribio16}, respectively (in this comparison, we used the same region as \citealt{toribio16}).

We exploit the simultaneous coverage of the H$\alpha$ and [NII] lines to derive oxygen abundances via the strong line method. As opposed to the direct (T$_{e}$-based) method, the strong line method relies on empirically-derived calibrations which can be used to derive abundances from emission line ratios.  Given that the two emission lines used for the N2 strong line ratio are close in wavelength (and therefore depend less on the reddening correction) and the ratio does not include high ionization potential lines (e.g.~[OIII], hence not introducing non-trivial dependencies on the degree of ionization), here we use the N2 line ratio,

\begin{equation}\label{eq:o3n2}
\mathrm{N2 = log\bigg(\frac{[NII]\lambda6584}{H\alpha}\bigg)}
\end{equation}

\noindent and adopt the empirical calibration for this line ratio given by \citet{marino13}.  We find oxygen abundances in the range $8.32 < {\rm 12+log(O/H)} < 8.40$, with the derived value for 118A being in excellent agreement with the ($T_{\rm e}$-derived) value of $8.38\pm0.06$ reported in \citet{toribio16} (12 + log(O/H) = 8.36$\pm$ 0.08 vs.~8.38$\pm$0.06 for R23 in this work and \citeauthor{toribio16}, respectively).  

To derive HII region expansion velocities, we perform Gaussian fits (see Table \ref{tab:linefits}) to a total of six nebular emission lines in the wavelength interval $6500 \AA < \lambda < 6800 \AA$, which covers the H$\alpha$ line, as well as [NII]$\lambda\lambda$6548,84, [SII]$\lambda\lambda$6717,31, and He I $\lambda$6678.  The selection of these lines to derive ionized gas kinematics is motivated by the fact that (a) they are among the brightest nebular emission lines; (b) they lie close in wavelength, therefore reducing possible uncertainties introduced by sub-optimal wavelength calibrations; (c) with similar ionization energies, one can assume that these are originating from the same line-emitting gas \citep{m16}.  From the difference between the mean systemic velocities of the two regions of ${\sim}102$ and ${\sim}105$ km~s$^{-1}$ (for 118B and 119C, respectively, derived from VLA HI data, \citealt{faesi14}) and the wavelength shifts obtained from the fitted line centroids, we find expansion velocities of the order of a few times the isothermal sound speed. This is in agreement with the expansion velocities measured independently by \citet{kruijssen19}, who obtain a typical HII region expansion velocity of $9.4^{+0.8}_{-0.7}$~km~s$^{-1}$ by dividing the inferred GMC radii (taken to be the Gaussian dispersion of $14.8^{+1.0}_{-0.9}$~pc) by the time it takes stellar feedback to disperse the parent GMC ($1.5\pm0.2$~Myr).

\begin{deluxetable*}{lccCcccc}
\tablecaption{Physical parameters of the five HII regions. \label{tab:strucprop}}
\tablecolumns{7}
\tablenum{5}
\tablewidth{0pt}
\tablehead{
\colhead{Region} & \colhead{$R_{90}$} & \colhead{$A_{V}$} & \colhead{$n_{\rm e}$} & \colhead{$T_{\rm e}$} & \colhead{12+log(O/H)} & \colhead{$v_{\rm exp}$} & \colhead{$M_{\rm ion}$}\\
\colhead{} &  \colhead{(pc)} & \colhead{mag} & \colhead{(cm$^{-3}$)} &  \colhead{($10^3$~K)} & \colhead{} & \colhead{(km s$^{-1}$)} & \colhead{(M$_{\odot}$)} 
}
\startdata
118A & 32.5 & 0.65 & 62.3$\pm$8.3 & 9.7$\pm$0.3 & 8.32$\pm$0.01 & 17.3$\pm$1.0 & 5.4$\times$10$^{3}$\\
118B & 25.7 & 0.71 & 41.4$\pm$12.7 & 11.2$\pm$0.9 & 8.39$\pm$0.01 & 27.6$\pm$2.6 & 1.0$\times$10$^{3}$\\
\hline
119A & 42.2 & 0.41 & 84.4$\pm$2.8 & 9.2$\pm$0.1 & 8.36$\pm$0.01 & 12.7$\pm$2.1 & 7.0$\times$10$^{3}$\\
119B & 27.6 & 0.37 & 55.9$\pm$3.1 & 8.9$\pm$0.1 & 8.40$\pm$0.01 & 8.4$\pm$3.8 & 2.3$\times$10$^{3}$\\
119C & 28.6 & 0.48 & 75.9$\pm$9.0 & 9.5$\pm$0.3 & 8.38$\pm$0.01 & 11.3$\pm$7.1 & 0.9$\times$10$^{3}$\\
\enddata
\end{deluxetable*}

\subsection{Luminosities, gas masses and feedback quantities}\label{sec:lum}
As in \citetalias{mcleod18b}, we derive HII region luminosities, $L_{H\alpha}$, from the (extinction-corrected) H$\alpha$ line intensity, and (under the assumption of Case~B recombination) convert these to Lyman continuum fluxes, $Q_{\rm H\alpha} = 7.31 \times L_{\rm H\alpha}$~s$^{-1}$ \citep{osterbrock}.  We also compute the total ionizing flux, $Q_{0,*}$, for each region as derived from the massive stars within them (all values are reported in Table \ref{tab:intprop}).  For this we assume that the stars within $R_{90}$ are spatially coincident and part of the respective regions, i.e.\ 118A contains 118-1, 118-3 and 118WR-1, 118B contains the star 118-4, 119A encompasses 119-2, 3, 4, 5, 6, 8 and 9, 119-7 and 119WR-1 are in the region 119B, and 119C is being inflated by 119WR-2. 

\begin{figure}[ht!]
\epsscale{1.2}
\plotone{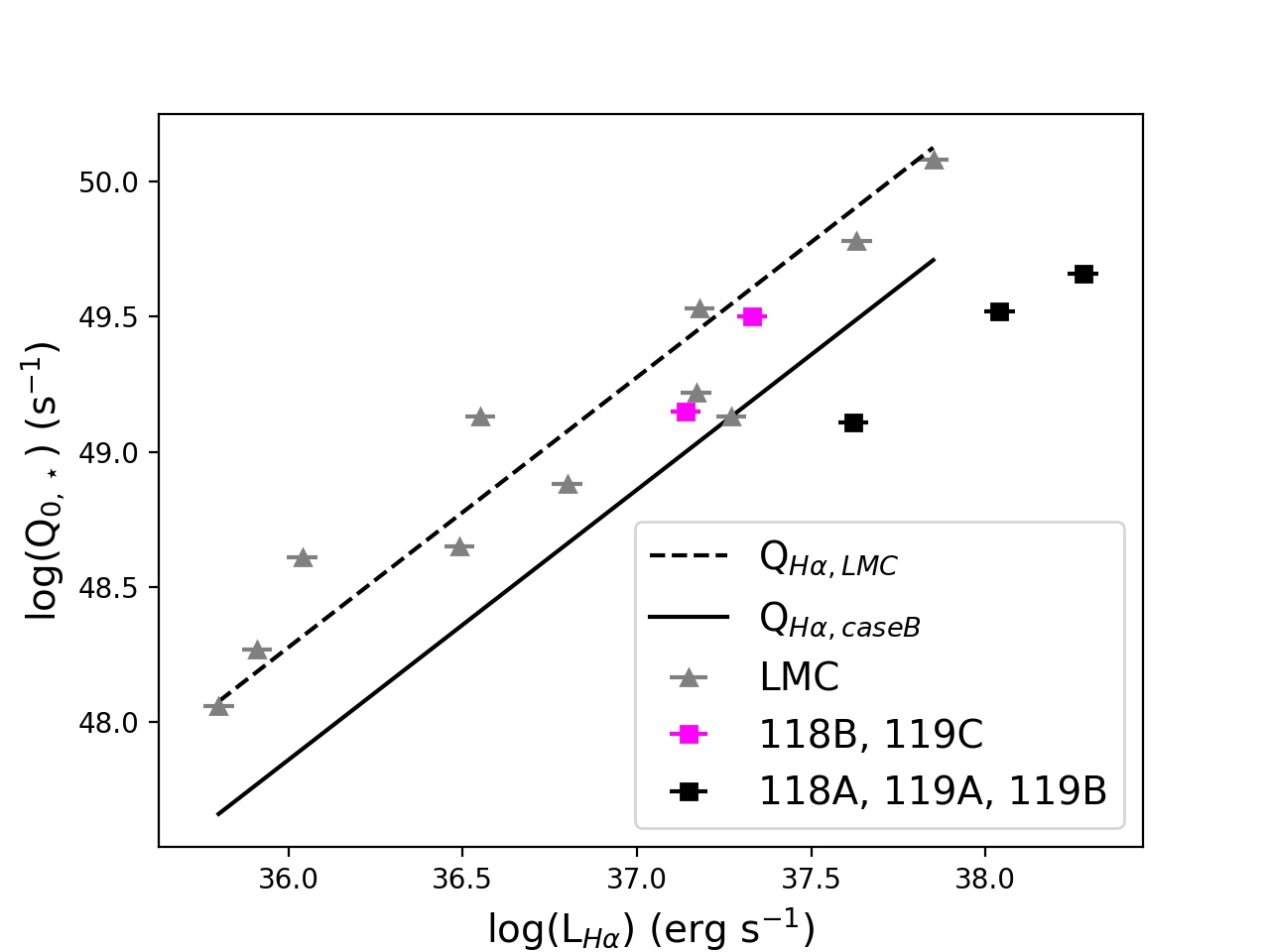}
\caption{Comparison between the photon flux $Q_{0,*}$ derived from the identification of the massive stars and as derived from Case~B recombination (solid black line).  Gray data points are HII regions in the LMC from \citetalias{mcleod18b}, and the dashed line is a fit to the LMC values.  Magenta squares are the two regions 118B and 119C, black squares are the remaining three (see Section~\ref{sec:lum}).}
\label{fig:lhaq}
\end{figure}

When compared to the photon flux $Q_{0,*}$ derived from the identification of the massive stars, we find that for the Case~B-derived photon flux $Q_{\rm H\alpha}$, only 118B and 119C are consistent with the trend found in the LMC (see Fig.~\ref{fig:lhaq}), i.e.\ not all of the ionizing photons emitted by the massive stars in these two regions go into producing the observed H$\alpha$ luminosity, and hence $Q_{\rm H\alpha} < Q_{0,*}$. For the remaining three HII regions, however, the opposite is true, i.e.~the derived photon flux cannot account for the observed H$\alpha$ luminosity ($Q_{\rm H\alpha} > Q_{0,*}$).  Our completeness in terms of the massive (O-type) stellar population is most likely significantly better in the analyzed LMC HII regions.  Both 118B and 119C contain a single identified massive star (118-4 and 119WR-2, respectively), and the opposite trend observed for 118A, 119A, and 119B might be indicating that we are missing some fraction of massive stars in these three regions.  There are several possible reasons for this, which we briefly discuss.  

First, we note that there are a few candidate massive stars which could not be reliably classified as O-type stars due to infilling of the He II line. Second, even with the high spatial resolution HST catalog as a reference, crowding remains challenging (particularly  in regions 118A and 119A), and higher spatial resolution IFU follow-up observations would be required.  The MUSE data discussed here were taken prior to MUSE having adaptive optics (AO) compatibility in the wide field mode, which is now available.  For example, targeted MUSE+AO observations in the narrow field mode would allow an order of magnitude better angular resolution, with the capability of reaching ${\sim} 0.3$~pc resolution, which is roughly the resolution we achieved for LMC HII regions and a factor $\sim$ 3 closer to the spatial resolution of HST/ACS. Third, some of these stars are likely to be unresolved binaries \citep[e.g.][]{sana12}, and as a consequence we are underestimating the emitted photon flux.  Lastly, there is likely a population of embedded O stars, which are therefore missed.  

We crudely investigate the (in)completeness of our identification algorithm by injecting O stars into a simulated MUSE cube, as described in Section \ref{sec:complete}.  With this procedure we find that we are systematically missing $\approx$ 30\% of the O star population, where the unidentified stars are those with low He II absorption line amplitudes, rather than those in crowded environments.  In terms of our observations, this translates into stars not being identified because they are either embedded or faint, low-mass stars.  This also shows the unique capability of the combination of high spatial resolution HST photometry and MUSE spectroscopy in identifying stars in crowded fields at 2 Mpc.

The combination of unresolved binaries and fainter or embedded stars being missed results in the derived photon fluxes Q$_{0,*}$ being strict lower limits, in particular for three out of the five regions (118A, 119A, 119B).  However, assuming that the two single sources in 118B and 119C are the main feedback-driving sources given the lack of other bright stars within them in both the HST and the MUSE data (hence assuming a complete census of massive stars within these specific regions), and assuming that the deviation from Case~B recombination conversion between H$\alpha$ luminosity and ionizing photon flux is due to photon leakage, we can compute the photon escape fraction, $f_{\rm esc} = (Q_{0,*} - Q_{\rm H\alpha}) / Q_{0,*}$, yielding $\sim 0.28\pm0.06$ and $\sim 0.51\pm0.10$ for 118B and 119C, respectively.  These values are consistent with analytical predictions for the later stages of HII region expansion \citep{rahner17} and recent measurements of the diffuse ionized gas fraction in nearby galaxies \citep[ranging from 0.25--0.75, see e.g.][]{kreckel18,kruijssen19,chevance19,hygate19b}.  As discussed in \citet{niederhofer16}, escaping photons are not absorbed in the HII region matter, but will instead contribute to the overall energetics of the host galaxy, consistent with the close correspondence between the escape fractions estimated for individual HII regions here and the global diffuse ionized gas fractions from the literature quoted above.

To analyze and quantify the feedback effect from massive stars, we proceed in computing feedback-related pressure terms, i.e.~the direct radiation pressure, $P_{\rm dir}$, the pressure of the warm, ionized gas, $P_{\rm ion}$, and the pressure due to stellar winds, $P_{\rm w}$, and are listed in Table~\ref{tab:intprop}. 

The direct radiation pressure is evaluated from the combined (bolometric) luminosity $L_{\star}$ of the massive stellar population in each region,
\begin{equation}
P_{\mathrm{dir}} = \frac{3L_{\star}}{4\pi R^{2}c}
\label{pdir}
\end{equation}
This is different from the expression used in \citetalias{mcleod18b}, which estimates the radiation force density at the rim of a shell rather than the volume-averaged radiation pressure. The pressure of the ionized gas is evaluated from the electron density as
\begin{equation}
P_{\mathrm{ion}}=(n_{\mathrm{e}}+n_{\mathrm{H}}+n_{\mathrm{He}})kT_{\mathrm{e}}\approx2n_{\mathrm{e}}kT_{\mathrm{e}}
\label{pion}
\end{equation}
where we assume an HII region temperature of 10$^{4}$ K as well as singly-ionized helium. The pressure from stellar winds follows from 
\begin{equation}
\begin{aligned}
P_{w}&\simeq 2.3\times10^{-12}\Big(\frac{L_{w}}{10^{36}\hspace{0.1cm}\mathrm{erg}\hspace{0.1cm}s^{-1}}\Big)^{2/5}\Big(\frac{\mathrm{n_{0}}}{0.25\hspace{0.1cm}\mathrm{cm}^{-3}}\Big)^{3/5}\\
&\Big(\frac{10^{6}\hspace{0.1cm}\mathrm{yr}}{t}\Big)^{4/5}\hspace{0.5cm}\mathrm{dyn}\hspace{0.1cm}\mathrm{cm^{-2}}
\end{aligned}
\label{pw}
\end{equation} 
with L$_{w}$ the wind luminosity, R = R$_{90}$, $n_{0}$ = $n_{\mathrm{e}}$/0.7 \citep{pillars} and $t$ the dynamical timescale \citepalias{mcleod18b}.

As already found for HII regions in the LMC and SMC \citep{lopez14,mcleod18b}, the direct radiation pressure is about one to two orders of magnitude lower than the other two pressure terms.  Hence, at these evolved HII region stages, the direct radiation pressure contributes to, but never dominates the expansion of the regions, which is driven by the combination of stellar winds and the pressure of the ionized gas. This matches the conclusion reached by \citet{kruijssen19}, who used an independent methodology complementary to ours to investigate the dominant feedback mechanisms driving GMC dispersal across a large fraction of the optical disk of NGC 300.  

In terms of $P_{\rm{dir}}$, we note that the derived values do not account for photon leakage, and likely don't represent a fully sampled massive end of the stellar population (see above).  This is particularly true for regions 118A, 119A and 119B.  However, even by assuming that the ionizing photons emitted by a {\it complete} stellar population are producing the observed H$\alpha$ luminosity and accounting for escaping photons, and hence using $Q_{\rm H\alpha}$ instead of $Q_{0,*}$, $P_{\rm{dir}}$, would still be one to two orders of magnitude lower than the other two pressure terms.

Furthermore, we compute the mass of the ionized gas according to \citet{osterbrock},

\begin{equation}\label{eq:mion}
    M_{\rm ion} \approx Q_{L({\rm H\alpha})} m_{\rm p} / (n_{\rm e}\alpha_{\rm B})
\end{equation}

\noindent with the proton mass $m_{\rm p}$ and the Case~B coefficient $\alpha_{\rm B}$, and assuming negligible dust absorption.  The inferred ionized gas masses (listed in Table~\ref{tab:strucprop}) will be used in the following section to derive ionized gas mass loading factors.

\begin{deluxetable*}{lcccCcccccc}
\tablecaption{Integrated HII region properties. Column~2 corresponds to the total ionizing photon flux emitted by the massive stars within each region; column~3 is the integrated H$\alpha$ luminosity; column~4 corresponds to the photon flux as derived from $L_{\rm H\alpha}$; columns~5 and 6 are the star formation rate and star formation rate density, respectively; the feedback related pressure terms $P_{\rm{dir}}$, $P_{\rm{ion}}$ and $P_{\rm{w}}$ (in units of $10^{-10}$ dyn cm$^{-2}$) are listed in colums~7, 8, and 9; column~10 lists the ionized gas mass loading factor.  See Appendix \ref{sec:errors} for uncertainties on the obtained quantities.  \label{tab:intprop}}
\tablecolumns{11}
\tablenum{6}
\tablewidth{0pt}
\tablehead{
\colhead{Region} & \colhead{log $Q_{0,\star}$} & \colhead{log $L_{0,\star}$} & \colhead{log $Q_{\rm H\alpha}$} & \colhead{log $L_{\rm H\alpha}$} & \colhead{SFR} & \colhead{$\Sigma_{\rm SFR}$} & \colhead{$P_{\rm{dir}}$} & \colhead{$P_{\rm{ion}}$} & \colhead{$P_{\rm{w}}$} & \colhead{$\eta_{\rm ion}$} \\
\colhead{} & \colhead{(s$^{-1}$)} & \colhead{(erg s$^{-1}$)} &\colhead{(s$^{-1}$)} & \colhead{(erg s$^{-1}$)} & \colhead{($10^{-4}$ M$_{\odot}$ yr$^{-1}$)} & \colhead{(M$_{\odot}$ yr$^{-1}$ pc$^{-2}$)} & \colhead{} & \colhead{} & \colhead{}
& \colhead{}}
\startdata
118A & 49.52 & 5.7 & 49.90 & 38.04 & 5.8 & 8.7$\times$10$^{-8}$ & 0.015 & 1.67 & 1.92 & 5.0\\
118B & 49.15 & 5.9 & 49.01 & 37.14  & 0.8 & 1.9$\times$10$^{-8}$ & 0.038 & 1.28 & 3.15 & 15.0\\
\hline
119A & 49.66 & 6.0 & 50.15 & 38.28 & 10.4 & 9.3$\times$10$^{-8}$ & 0.018 & 2.15 & 1.38 & 2.1\\
119B & 49.11 & 5.3 & 49.49 & 37.62 & 2.3 & 4.8$\times$10$^{-8}$ & 0.0084 & 1.37 & 0.40 & 3.2\\
119C & 49.50 & 5.6 & 49.19 & 37.33 & 1.1 & 2.1$\times$10$^{-8}$ & 0.016 & 1.98 & 0.98 & 3.0\\
\enddata
\end{deluxetable*}

\subsection{Star formation rates, ionized gas mass loading factors and feedback efficiency}\label{sec:sfr}

From $L_{\rm H\alpha}$ we derive star formation rates (SFRs) according to \citet{kennicutt12},

\begin{equation}
\mathit{\rm log(SFR)}~({\rm M}_{\odot}~{\rm yr}^{-1}) = {\rm log}~L_{\rm H\alpha}~-~{\rm log}~C_{\rm H\alpha}
\label{sfr}
\end{equation}

\noindent (with log $C_{\rm H\alpha} = 41.27$, \citealt{hao11}) and, together with $R_{90}$, the star formation rate surface density, $\Sigma_{\rm SFR}$.  We note that using the above $L_{\rm{H\alpha}}$-SFR conversion on spatially-resolved regions such as the ones studied here introduces a series of caveats, as this conversion (i) is based on an ensemble average over HII regions of different ages; (ii) it was derived for solar abundances; (iii) it relies on population synthesis models.  Indeed, at SFRs below $\sim~0.001-0.01$ M$_{\odot}$~yr$^{-1}$, stellar populations in the regions do not fully sample the stellar initial mass function (IMF), leading to the luminosity of the used tracer to fluctuate on small spatial scales for a given SFR \citep{kennicutt12}.  Tracers of the ionized gas, e.g.~the H$\alpha$ luminosity used here, are particularly affected by this, given that these mostly trace the upper end of the IMF. However, under the assumption of the gas and stars in the analyzed regions being physically associated (i.e.~the regions being young), deriving SFRs remains a meaningful exercise to not only estimate the level of star formation activity in the youngest regions across NGC 300 and to compare these values to those derived for the LMC HII regions, but also to evaluate the reliability of using common galaxy-scale analysis methods and relations for cloud-scale studies.

Due to the overall larger sizes and higher luminosities compared to the LMC HII region complexes N44 and N180 \citepalias{mcleod18b}, we find overall higher SFRs in these NGC 300 HII regions.  Despite these relatively high star formation rates, none of these compare to e.g.\ 30~Doradus, one of the most massive star-forming regions in the nearby Universe, where SFRs of the order of some $0.02$ M$_{\odot}$~yr$^{-1}$ over radii of ${\sim}140$~pc have been measured \citep{ochsendorf17}.

\begin{figure}[ht!]
\gridline{\fig{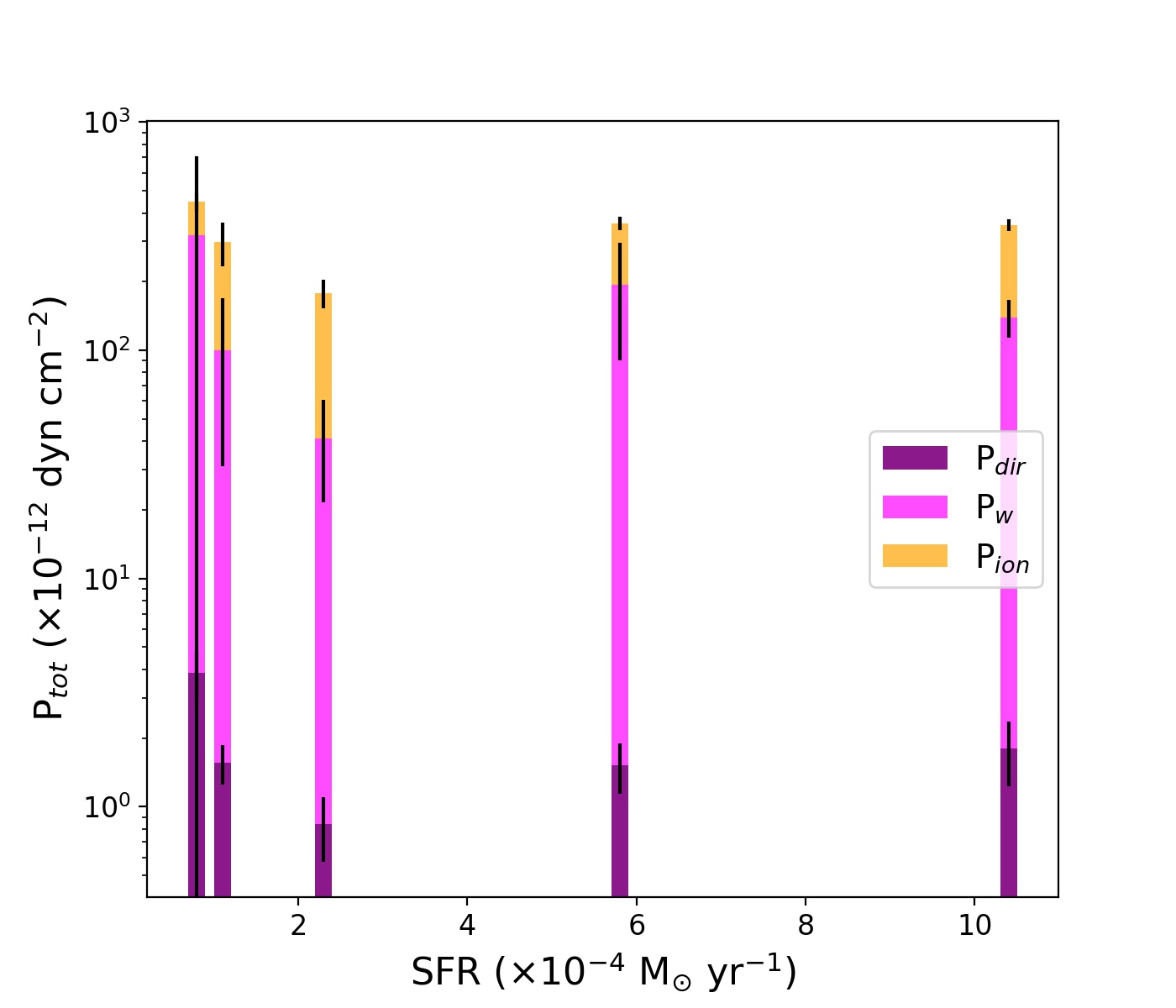}{0.5\textwidth}{(a)}}
\gridline{\fig{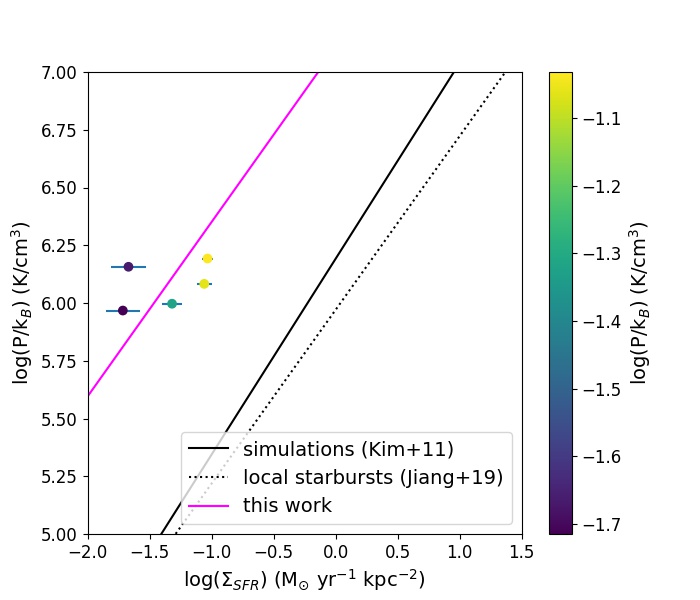}{0.5\textwidth}{(b)}}
\caption{(a) Relative contribution of the various pressure terms to the total pressure as a function of the star formation rate. (b) Relation between the SFR surface density and the ionized gas pressure. To compare our findings with simulations \citep{kim11} and observations \citep{jiang19} of local analogs of high-redshift galaxies (dashed and dotted lines, respectively), here $P_{\rm ion}$ is in units of K cm$^{-3}$ and $\Sigma_{\rm SFR}$ in units of M$_{\odot}$ yr$^{-1}$ kpc$^{-2}$. The solid line shows the simulated relation shifted down by a factor of 30.}
\label{fig:bar}
\end{figure}

Fig.~\ref{fig:bar}(a) shows the relative contribution of the various pressure terms to the total pressure as a function of SFR.  As found for the LMC regions, the pressure of the ionized gas and the pressure from stellar winds contribute almost equally to the total pressure.  An exception to this is 118B, for which we derive the, overall, lowest SFR and highest ionized gas mass loading factor (see below), together with the highest stellar wind pressure term.  With 118B being the smallest of the five regions, it also has the highest measured expansion velocity.  This region is associated with a single O-type star, the high-mass object [DCL88]118-4, and the measured H$\alpha$ luminosity of this ``single-star'' bubble is consistent with the low star formation rate and strong stellar wind. 

Overall, we find a weak trend between the ionized gas pressure and $\Sigma_{\rm SFR}$, however we refrain from drawing any conclusions for the NGC 300 HII regions due to the small number of data points.  In this sense, the linear fit in Fig.~\ref{fig:bar}(b) is only indicative and associated with a large uncertainty on the slope of the fitted line, and needs to be revised with the analysis of the full MUSE data set across NGC 300.  This trend is observed throughout cosmic time from e.g. $z = 2.5$ H$\alpha$ emitters \citep{shimakawa15} to local analogs of these \citep{jiang19}, as well as in simulations \citep{kim11}.  Fig.~\ref{fig:bar}(b) shows our sample of 5 HII regions together with the relations derived for nearby compact starbursts \citep{jiang19} and simulations \citep{kim11}.  It is widely accepted that on the scales of small, individual star-forming regions, other galaxy-wide scaling relations (e.g.\ the Kennicutt-Schmidt relation) are no longer valid and, as discussed in \citet{kl14} (see also Section \ref{sec:outlook}), the main reason for these galaxy-wide scaling relations breaking down on small scales is that these trace time-dependent phenomena such as the conversion of gas into stars, of which single regions are mere snapshots.  One would therefore conclude that trying to look for such relations for individual regions is not particularly insightful. 

However, there are cases in which properties derived by integrating over entire galaxies can be overestimated.  This is the case of the $P_{\rm ion} - \Sigma_{\rm SFR}$ relation: as discussed in \citet{kewley19}, electron densities and gas pressures derived from integrated galaxy spectra are contaminated by the diffuse ionized gas (DIG, caused by HII region photon leakage, shock-excitation and dust-reprocessed radiation).  This can result in overestimates of these quantities, which makes a small-scale ($<$ 100 pc) investigation of the $P_{\rm ion} - \Sigma_{\rm SFR}$ relation meaningful and necessary. Indeed, the location of the NGC 300 HII regions in the parameters space shown in Fig.~\ref{fig:bar}(b) is consistent with the above argument.


We combine the mass and kinematics of the ionized gas with the SFR to derive the ionized gas mass loading factors (i.e.\ the mass outflow rate of the ionized gas due to stellar feedback per unit SFR) for the five HII regions,

\begin{equation}
    \eta_{\rm ion} = \frac{\dot{M}_{\rm ion}}{\mathit{SFR}}
\end{equation}

\noindent where $\dot{M}_{\rm ion}$ = $M_{\rm ion}v/R_{90}$ is the ionized gas mass outflow rate.  Stellar feedback in these regions is dominated by stellar winds and the ionized gas pressure, which therefore are driving the mass removal quantified by $\eta_{\rm ion}$.  For all five regions we find $\eta_{\rm ion} > 1$, implying a negative feedback effect.  The combined effect of stellar winds and ionized gas pressure is therefore efficiently quenching star formation by clearing the gas and driving the expansion of the HII regions. 

\citet{kruijssen19} recently combined 20-pc resolution maps of molecular gas and H$\alpha$ emission in NGC 300 to derive a cloud-scale total gas mass loading factor of $\eta_{\rm tot}=40^{+31}_{-18}$ across the star-forming disk of NGC 300. This mass loading factor exceeds our values by about an order of magnitude, which would suggest that the ionized gas mass does not dominate the mass budget of the expanding bubbles. Indeed, both D118 and D119 are associated with nearly $10^5$~M$_\odot$ of molecular gas \citep{kruijssen19}, which is roughly an order of magnitude larger than our ionized gas masses. Therefore, the difference in mass loading factors may either be a direct result of the multi-phase nature of the feedback-driven bubbles, or implies that these two HII region complexes are particularly young and still contain a large molecular gas reservoir.

The range of values for $\eta_{\rm ion}$ derived for the five NGC 300 HII regions are larger than those found for star-forming clumps in low-$z$ luminous star-forming galaxies \citep{arribas14} where $\eta_{\rm ion}$ is generally below unity and no negative feedback is observed.  We note that while they are not entire galaxies, the regions analyzed in \citet{arribas14} have kpc-scale sizes.  These apertures are therefore likely to contain populations of unresolved HII regions as well as DIG, the latter also contributing to the measured H$\alpha$ luminosity and therefore to the derived SFRs, resulting in lower ionized gas mass loading factors.  Galaxy-wide IFU studies that resolve individual HII regions will therefore deliver ground-breaking progress thanks to the ability of being able to efficiently remove large-scale contaminating factors to galaxy scaling relations.

\section{Outlook}\label{sec:outlook}
The data cubes used in this work are a mere $2$ of a 35-pointing mosaic that covers most of the star-forming disk of NGC 300. The full MUSE NGC 300 data set samples ${>}100$ individual HII regions, as well as a large number of planetary nebulae, supernova remnants, red supergiants, and X-ray binaries.  In McLeod et al.\ (in preparation) we will analyze the full NGC 300 data set in terms of massive stars, stellar feedback, and HII region properties.  In this forthcoming publication we will not only deliver all the quantities derived in here for the entire HII region population in our data set, hence producing comprehensive catalogs of massive stars, gas kinemeatics, physical HII region parameters and feedback quantities, we will also publish the reduced data cubes for the community to use. 

Together with the NGC 300 catalogs and data set we will also publish the analysis methods and techniques used in this paper as {\sc Python} Notebooks, which will be readily applicable to generic IFU data sets.  Below, we briefly discuss some of the more immediate deliverables of the full MUSE NGC 300 data. 

\subsection{Finding Wolf-Rayet stars}\label{sec:findwr}
In terms of stellar feedback analyses, WR stars are among the dominant feedback sources due to their strong ionizing spectra, powerful stellar winds, and due to the fact that most of these are supernova progenitors.  In this context, the decrease in strength of stellar winds at lower metallicities is particularly interesting.  As opposed to the inner regions of the Magellanic Clouds which have a rather shallow metallicity gradient \citep{feast10,choudhury18}, the star-forming disk of NGC 300 has a steeper gradient \citep[e.g.][]{toribio16}, and we would therefore expect weaker WR star winds in the outer part of the disk.  Moreover, because the ratio of WC to WN stars is shown to be sensitive to metallicity \citep{massey96}, we would also expect to find more WN stars in the outer parts of NGC 300.  Hence, identifying WR stars is essential when studying the feedback energy budget of HII regions throughout galaxies.  

The WR content of NGC 300 has been subject to several past studies \citep[e.g.][]{breysacher97,schild03}, and the number of known WR stars in the galaxy is currently at ${\sim}60$ sources.  These studies are predominantly based on narrow-band imaging, combined with spectroscopic follow-up for selected sources.  Because MUSE covers the entire optical regime (i.e.~4750 - 9350 \AA) and we are able to make images of any narrow- or broad-band filter within the instrument's wavelength coverage, we are able to simultaneously photometrically identify and spectroscopically confirm  WC, WO and WN stars with the same data set.  In McLeod et al.\ (in preparation) we will apply this to the entire $7\arcmin{\times}5\arcmin$~ MUSE mosaic of NGC 300.  Because we are also able to retrieve independent metallicity measurements from the HII regions, the planetary nebulae, and the red supergiants in the mosaic, we will be able to directly analyze the contribution of WR winds and radiation to the feedback energy budget of the individual HII regions as a function of metallicity. 

\subsection{Connecting feedback to integrated HII region properties across the NGC 300 disk}
In the context of the dependence of stellar feedback on the properties of the regions the massive stars form in, i.e.\ gas-phase metallicities, gas densities, star formation history (e.g.~\citealt{gogarten10}), location within the galaxy, it is imperative to obtain HII region properties and massive stellar censuses for a statistically significant sample, rather than only for a handful of well-studied regions. 

As already introduced in the previous section, large IFU surveys of entire galaxies allow, for example, metal abundance determinations from hundreds of objects representing different evolutionary stages or epochs of star formation {\it simultaneously}, e.g.\ HII regions (e.g.\ \citealt{toribio16}), planetary nebulae (e.g.\ \citealt{stasinska13}), red supergiants (e.g.\ \citealt{gazak15}).  By connecting WC/WN populations to gas-phase metal abundances (e.g.\ C/O, N/O), this will enable to directly relate feedback to the chemical enrichment and dispersal history.  

Of particular interest will be the ability of comparing the derived properties of hundreds of HII regions to feedback models, e.g. {\sc warpfield} \citep{rahner19}.  Specifically, because the observations sample the timeline of star formation, such comparisons will enable the analysis of time-dependent HII region diagnostics.  A first comparison between these and SITELLE IFU observations of NGC 628 \citep{rousseau18} is shown in \citet{pellegrini19}, which demonstrates the power to harness the observations to test the models, and to harness the models to understand the observations.

\subsection{The multi-phase interstellar medium and the small-scale physics of star formation and feedback}
As discussed in \citet{kl14}, scaling relations like the Kennicutt-Schmidt relation break down on the small scales of individual GMCs and HII regions.  Galactic scaling relations are built from integrated observations of regions that are large enough to fully sample the complete evolutionary timeline from cloud formation to star formation and feedback-driven dispersal -- the small-scale breakdown is due to individual regions representing single evolutionary snapshots of this timeline rather than an ensemble average.

\citet{kruijssen18} developed a statistical method for translating this statistical behaviour of the relation between molecular gas (e.g.\ traced by CO) and star formation (e.g.\ traced by H$\alpha$) to empirical constraints on baryon cycle in the ISM down to GMC scales. This methodology provides direct measurements of the molecular cloud lifetime, the feedback timescale (i.e.\ the time for which GMCs and HII regions coexist), the feedback dispersal velocity, the cloud-scale total mass loading factor, and the integrated star formation efficiency per star formation event. This method has previously been applied to NGC 300 using CO and narrow-band H$\alpha$ data \citet[also see the comparisons to our results throughout Section~\ref{sec:gas}]{kruijssen19}, as well as about a dozen other nearby galaxies \citep{chevance19,hygate19b}.

While the MUSE coverage across a sizable part of NGC 300's star-forming disc is ideally suited for further applications of this methodology, of particular interest is the fact that any tracer probing part of the evolutionary timeline of GMCs and HII regions can be used to construct a multi-tracer timeline of star formation and stellar feedback. \citet{ward19} provide a recent example, measuring the HI cloud lifetime in the LMC. With the MUSE data obtained here, we will be able to measure the emission lifetimes of a wide variety of ionized emission lines, in conjunction with the physical quantities derived already by combining CO and H$\alpha$ emission.

\subsection{The diffuse ionized gas}\label{sec:dig}
As discussed in \citet{kewley19} and in Section \ref{sec:sfr}, small-scale ($<$ 100 pc) observations are the missing piece to disentangle the contribution of contaminating factors to galaxy-wide quantities and relations between these.  A prominent example of this is the effect of the diffuse ionized gas on measurements of e.g. abundance gradients, ionized gas pressures, or SFRs, which can be grossly overestimated.  Moreover, the diffuse ionized gas contamination is scale-dependent, and spatially-resolved observations of individual HII regions across entire galaxies are therefore vital in better constraining the quantities that are used to derive common galaxy-wide relations. The wide-field coverage of the full MUSE mosaic is ideally suited for combining statistical methods to isolate the diffuse ionized gas \citep[e.g.][]{hygate19} with a spatially-resolved, region-by-region census of individual HII regions. We plan to explore this possibility in future work.

\section{Summary and conclusions}\label{sec:end}
This article presents a pilot study of resolved stellar feedback in five HII regions in the nearby spiral galaxy NGC 300, covered with a $1\arcmin{\times}2\arcmin$ integral field unit data set from the MUSE instrument.  We demonstrate that at $2$~Mpc distance, in combination with high spatial resolution HST photometry, IFU data can be used to simultaneously identify the feedback-driving massive stars and obtain kinematics, physical properties and feedback-related quantities of the individual star-forming regions hosting the massive stars.  

We spectroscopically identify $13$ previously unknown O-type stars associated with the HII regions and derive atmospheric parameters. Moreover, we revisit the spectral type of two WR stars and confirm the WR nature of a further two previously known candidates.  The novelty of combining high angular resolution HST data with the ground-based IFU cubes demonstrates that at a distance of $2$~Mpc we are able to de-blend individual sources in crowded star-forming regions. 

By comparing the ionizing photon budget of each region as derived from its stellar content to that derived from the measured H$\alpha$ luminosity, we conclude that the massive star census is likely not fully complete for three of the five HII regions (118A, 119A and 119B).  The incompleteness is likely due to a combination of unresolved binaries and embedded or faint O-type stars, and the derived combined photon fluxes $Q_{0,*}$ of these regions are therefore strict lower limits.  For two of the five HII regions we can assume a complete census of massive, feedback-driving stars and for these we find that the derived fraction of photons escaping from the HII regions is 28~and 51\%, consistent with the later stages of HII region expansion.  If measured for a large number of HII regions across an entire galaxy, the escape fraction can in be compared to the diffuse H$\alpha$ emission of the NCG 300 disk to evaluate the amount of photons leaking from HII region and therefore going into the heating budget of the galaxy. \citet{kruijssen19} perform this measurement for the entire star-forming disk of NGC 300 and find a diffuse ionized gas fraction of 33\%, within the range spanned by the two HII regions for which we can currently calculate this number. Targeted follow-up IFU observations with higher angular resolution (e.g.\ AO-assisted MUSE narrow field mode observations) and combined with NIR spectroscopy will allow us to obtain a more complete census of massive stars. 

For the HII regions discussed here we derive ionized gas kinematics, compute feedback-related pressure terms, luminosities, star formation rates and star formation rate surface densities, ionized gas masses, ionized gas mass loading factors, oxygen abundances, electron densities and temperatures.  By comparing the obtained quantities to a similar study of HII regions in the LMC, we confirm that the dynamics of the HII regions is dominated by the warm ionized gas and stellar winds, while the direct radiation pressure is ${\sim}2$ orders of magnitude lower than the former two.  Moreover, we compare the ionized gas mass loading factors derived for the 5 HII regions to the {\it total} gas mass loading factor derived for the NGC 300 disk \citep{kruijssen19}, and find that the mass budget of the expanding bubbles is not dominated by the ionized gas.  This is either directly linked to the multi-phase nature of these regions, or indicating very early evolutionary stages and consequently higher molecular gas masses.

We furthermore take advantage of being able to derive the pressure of the ionized gas, enabling to evaluate the $P_{\rm ion} - \Sigma_{\rm SFR}$ relation without the contamination by the diffuse ionized gas which is problematic at large (kpc) scales where $P_{\rm ion}$ values are systematically overestimated.  Carried out over the full MUSE mosaic of NGC 300, this will, for the first time, allow us to confidently derive this relation (as well as other scaling relations) without the diffuse ionized gas contamination.

In terms of a broader picture of the role of stellar feedback in star-forming galaxies, we briefly discuss the implications and potential outcomes of applying the analysis techniques and methods derived in this proof-of-concept study on large IFU data sets covering entire galaxies.  A first study of this scale will be presented in a forthcoming publication encompassing the central $7\arcmin{\times}5\arcmin$ region covering most of the star-forming disk of NGC 300.  This short study therefore explores the necessary techniques for resolved stellar feedback studies in nearby galaxy scales. These will be readily applicable to data sets from ongoing and forthcoming nearby galaxy IFU surveys, and therefore deliver the needed statistics to observationally quantify stellar feedback over representative galactic properties and environments.

\acknowledgments
Support for this work was provided by NASA through the NASA Hubble Fellowship grant
HF2-51442.001 awarded by the Space Telescope Science Institute, which is operated by the
Association of Universities for Research in Astronomy, Inc., for NASA, under contract
NAS5-26555.  This research is partly supported by a Marsden Grant from the Royal Society of New Zealand (AFM). It is based on observations made with ESO Telescopes at the Paranal Observatory under program ID 098.B-0193. JMDK and MC gratefully acknowledge funding from the German Research Foundation (DFG) in the form of an Emmy Noether Research Group (grant number KR4801/1-1) and the DFG Sachbeihilfe (grant number KR4801/2-1). JMDK gratefully acknowledges funding from the European Research Council (ERC) under the European Union's Horizon 2020 research and innovation programme via the ERC Starting Grant MUSTANG (grant agreement number 714907) and Sonderforschungsbereich SFB 881 “The Milky Way System” (subproject B2) of the DFG.
AFM thanks M. Krumholz and C.~McKee for the useful discussion, and S.~Kamann for the invaluable help with PampelMuse.  PZ acknowledges support by the Forschungsstipendium (ZE 1159/1-1) of the German Research Foundation, particularly via the project 398719443. CMF acknowledges support from the National Science Foundation under Award No. 1903946.

%

\vspace{5mm}
\facilities{HST(ACS), VLT(MUSE)}
\software{MUSE pipeline \citep{weilbacher15}, PampelMuse \citep{kamann13}, PySpecKit \citep{pyspec}, PyNeb \citep{pyneb}, Astropy \citep{astropy:2018}.}

\appendix

\section{Appendix information}

\subsection{Completeness}\label{sec:complete}
As described in Section \ref{sec:lum}, we are most likely not detecting a fraction of the O stars in the clusters due to crowding, which at a distance of 2 Mpc is challenging even in the HST data. We therefore perform completeness tests to estimate the fraction of unidentified O-type stars. For this, we proceed as follows:

\begin{itemize}
    \item we simulate a $20\arcsec\times20\arcsec$ HST image (i.e.~$20\arcsec$ = $400$ pixel) by populating it with randomly-located Gaussian sources;
    \item with the simulated HST image we produce a data cube spanning from 5350.28 to 5469.03 \AA, i.e.~covering the He II$\lambda$5411 absorption line used to identify and classify the O stars, by propagating the simulated image along the spectral axis and adding random Gaussian noise to each frame; the spectral axis of the simulated data cube has the same sampling as the MUSE data (1.25 \AA);
    \item we produce a sample of synthetic O star spectra, thus having a He II$\lambda$5411 absorption line; for this we randomly sample line amplitudes and line widths, and we add noise to the spectra;
    \item next, we identify the (simulated) stars in the $97^{th}$ brightness percentile (motivated by the brightness of the O stars identified in the real data) and assign the pixels associated with these stars a spectrum as created above; all the other stars (because they are not O stars) will not have this absorption line;
    \item we then re-bin the synthetic HST data cube to the resolution of MUSE, i.e.~we obtain a synthetic MUSE data cube with $100\times100$ pixels (with a spatial sampling of 0.2\arcsec/pixel, this exactly corresponds to $20\arcsec\times20\arcsec$); pixels are averaged and combined;
    \item after re-binning, we further convolve the synthetic MUSE cube to a resolution of $\approx$ 1\arcsec, so to the seeing-limited resolution of Field 2 (see Table \ref{tab:pointings});
    \item finally, we proceed in extracting stellar spectra and identifying O stars as per Section \ref{sec:stars} based on the synthetic HST star catalog;
\end{itemize}

\begin{figure*}
\gridline{\fig{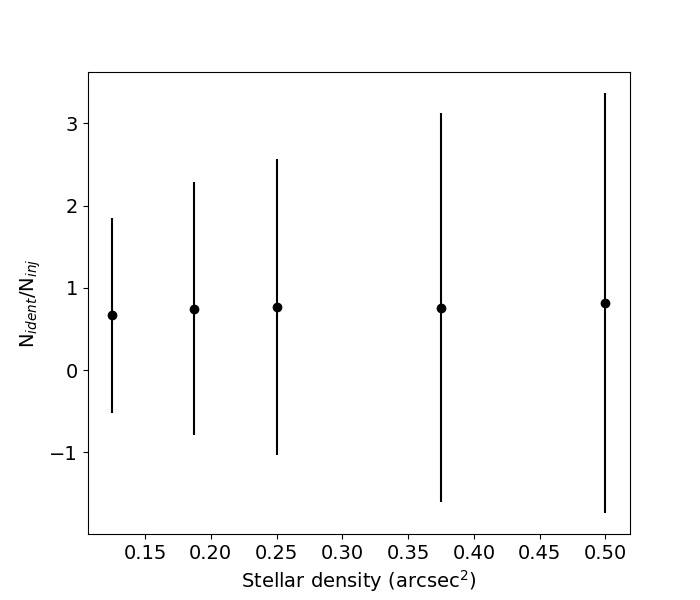}{0.5\textwidth}{(a)}
          \fig{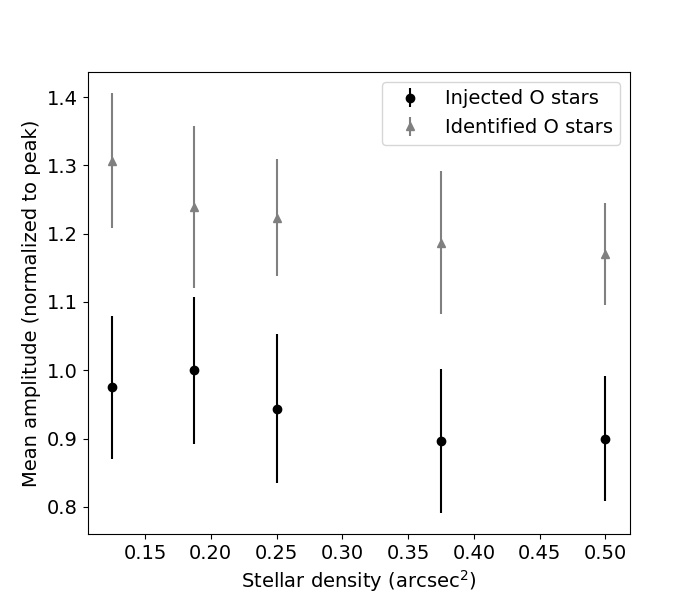}{0.5\textwidth}{(b)}
          }
\caption{(a) Ratio of the mean number of identified to injected O stars in the simulated MUSE cube as a function of stellar density, as described in \ref{sec:complete}, errors correspond to the standard deviation of the number of identified stars per bootstrap. (b) Mean amplitude of the He II$\lambda$ absorption line of the injected and identified O stars, showing that the sample of identified O stars is, on average, more luminous due to fainter stars being unidentified. \label{fig:boot}}
\end{figure*}

The above procedure is bootstrapped 100 times to account for random sampling statistics, and repeat the above procedure for different stellar densities to assess the completeness of identified O stars. This is shown in Fig.~\ref{fig:boot}, which shows the average number of identified stars as a function of stellar density (with $\sim$ 0.16 stars/arcsec$^{2}$ being the mean average density based on the HST photometry and the limiting magnitude of the MUSE data), as well as the mean amplitude of the HeII absorption line (normalized to peak). 

Fig.~\ref{fig:boot} illustrates how at $\sim$ 0.7 the average fraction of identified O stars (i.e.~with respect to injected O stars) is roughly constant over the explored stellar density range, and that the main contributing factor to not identifying stars lies in their effective temperature: O stars with lower T$_{eff}$ (i.e.~shallower HeII absorption lines) are systematically identified less than brighter, hotter stars.

\subsection{Uncertainty estimation}\label{sec:errors}

Values reported in Tables \ref{tab:strucprop} and \ref{tab:intprop} are derived via Gaussian fitting to the integrated spectra of the 5 HII regions, i.e.~from the circular apertures shown in Fig.~\ref{fig:ostars}.  The uncertainties of the computed quantities therefore obtained by rigorously propagating the errors from the Gaussian fits to the various emission lines (see Table \ref{tab:linefits}) through the various steps in computing them, including uncertainties coming from the extinction correction. Given the small uncertainties on the distance to NGC 300 \citep{dalcanton09}, we do not include it in our calculations. For brevity we do not report the propagated uncertainties in Tables \ref{tab:strucprop} and \ref{tab:intprop}. An exception to this are the luminosity of the WR stars, the values of which are from the best fit PoWR grid model, and for which we assume an accommodating 20\% uncertainty (see text Section \ref{sec:stars}).  

\subsection{Determination of R$_{90}$}\label{sec:r90}

To determine R$_{90}$ we crop the integrated H$\alpha$map (by eye) to contain only the relevant, individual regions. These sub-maps are then used as input for the radial profile algorithm, which determines the radius containing 90\% of the H$\alpha$ flux.

\begin{figure*}[ht!]
\plotone{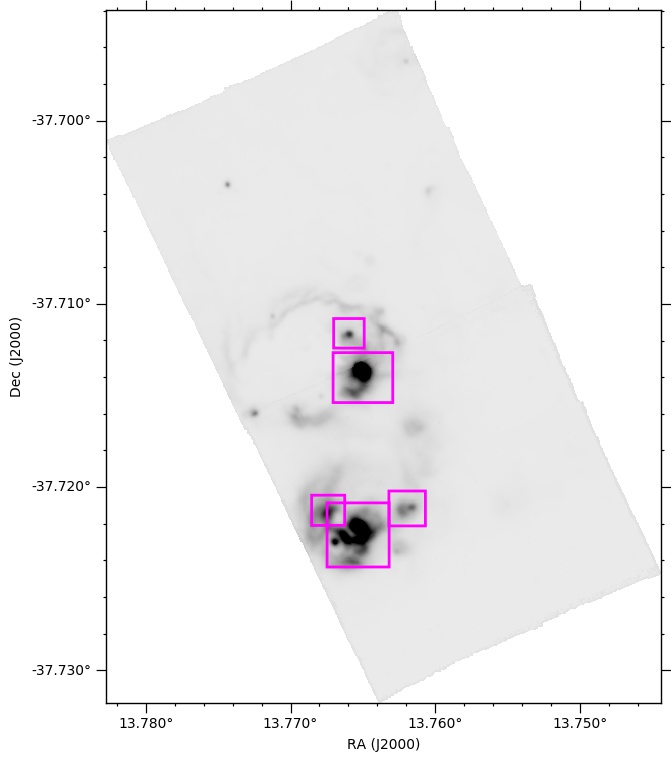}
\caption{H$\alpha$ map, magenta regions were used to determine R$_{90}$, defined as the radius (starting from the central coordinates of the respective regions) which encompasses 90\% of the H$\alpha$ emission.}
\label{fig:r90_fig}
\end{figure*}

\begin{figure}[ht!]
\epsscale{0.7}
\plotone{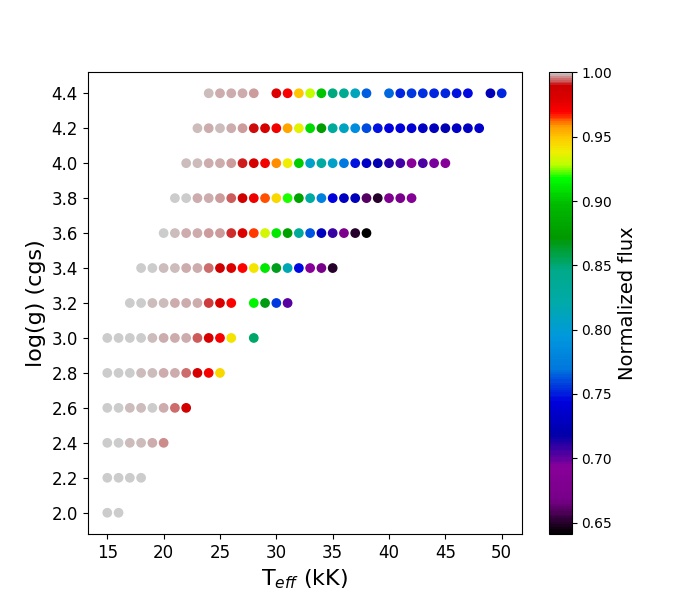}
\caption{Effective temperature versus surface gravity from the PoWR model grid \citep{hainich19}, color-coded by the depth of the He II$\lambda$5411 line of each synthetic spectrum.}
\label{fig:powr}
\end{figure}

\begin{deluxetable*}{lccCcccc}
\tablecaption{Gaussian line centroid fits (first row for each region) and centroid shifts with respect to the rest wavelength (second row for each region). All units are in \AA. \label{tab:linefits}}
\tablecolumns{8}
\tablenum{7}
\tablewidth{0pt}
\tablehead{
\colhead{Region} & \colhead{[NII]$\lambda$6548} & \colhead{H$\alpha$} & \colhead{[NII]$\lambda$6584} & \colhead{He I$\lambda$6678} & \colhead{[SII]$\lambda$6717} & \colhead{[SII]$\lambda$6731} & H$\beta$\\
}
\startdata
118A & 6550.68$\pm$0.16 & 6565.37$\pm$0.01 & 6586.05$\pm$0.05 & 6680.82$\pm$0.54 & 6719.17$\pm$0.07 & 6733.54$\pm$0.09 & 4863.24$\pm$0.02\\
 & 2.58$\pm$0.16 & 2.60$\pm$0.01 & 2.55$\pm$0.05 & 2.62$\pm$0.53 & 2.73$\pm$0.07 & 2.72$\pm$0.09 & 1.91$\pm$0.02\\
118B & 6550.90$\pm$0.36 & 6565.69$\pm$0.02 & 6586.29$\pm$0.12 & 6681.06$\pm$2.09$\pm$ & 6719.35$\pm$0.13 & 6733.71$\pm$0.18 & 4863.58$\pm$0.08\\
 & 2.80$\pm$0.36 & 2.92$\pm$0.02 & 2.79$\pm$0.12 & 2.86$\pm$2.09 & 2.91$\pm$0.13 & 2.89$\pm$0.18 & 2.25$\pm$0.08\\
\hline
119A & 6550.64$\pm$0.09 & 6565.35$\pm$0.01 & 6586.01$\pm$0.03 & 6719.14$\pm$0.04 & 6733.50$\pm$0.06 & 6680.81$\pm$0.44 & 4863.19$\pm$0.02\\
 & 2.54$\pm$0.09 & 2.58$\pm$0.01 & 2.51$\pm$0.03 & 2.61$\pm$0.44 & 2.70$\pm$0.04 & 2.68$\pm$0.06 & 1.86$\pm$0.02\\
119B & 6550.55$\pm$0.17 & 6565.25$\pm$0.01 & 6585.94$\pm$0.05 & 6680.66$\pm$0.99 & 6719.03$\pm$0.08 & 6733.39$\pm$0.11 & 4863.11$\pm$0.03\\
 & 2.45$\pm$0.17 & 2.48$\pm$0.01 & 2.44$\pm$0.05 & 2.46$\pm$0.99 & 2.59$\pm$0.08 & 2.57$\pm$0.11 & 1.78$\pm$0.03 \\
119C & 6550.60$\pm$0.35 & 6565.37$\pm$0.02 & 6585.99$\pm$0.11 & 6680.90$\pm$1.57 & 6719.08$\pm$0.13 & 6733.44$\pm$0.18 & 4863.21$\pm$0.06\\
 & 2.50$\pm$0.35 & 2.60$\pm$0.02 & 2.49$\pm$0.11 & 2.70$\pm$1.57 & 2.63$\pm$0.13 & 2.62$\pm$0.18 & 1.90$\pm$0.06\\
\enddata
\end{deluxetable*}




\bibliography{n300_refs}  




\end{document}